\documentclass[12pt,notitlepage,a4paper]{article}
\pdfoutput=1
\usepackage{color,graphicx}
\usepackage{cite}
\usepackage{amssymb}
\usepackage{delarray,amsmath,bbm}
\usepackage[latin1]{inputenc}
\usepackage[american]{babel}
\usepackage[table]{xcolor}
\usepackage{cite}
\usepackage{tikz-cd}

\newcommand{\be}{\begin{equation}}
\newcommand{\ee}{\end{equation}}
\newcommand{\beq}{\begin{equation}}
\newcommand{\eeq}{\end{equation}}
\newcommand{\beqa}{\begin{eqnarray}}
\newcommand{\eeqa}{\end{eqnarray}}

\newcommand{\bear}{\begin{eqnarray}}
\newcommand{\eear}{\end{eqnarray}}

\newcommand{\pd}{\partial}
\numberwithin{equation}{section}

\newfont{\namefont}{cmr10}
\newfont{\addfont}{cmti7 scaled 1440}
\newfont{\boldmathfont}{cmbx10}
\newfont{\headfontb}{cmbx10 scaled 1728}

\begin{document}
\baselineskip=15.5pt
\pagestyle{plain}
\setcounter{page}{1}

\begin{center}
\vspace{0.1in}

\renewcommand{\thefootnote}{\fnsymbol{footnote}}

\begin{center}
\Large \bf  Adiabatic  pumping solutions in global AdS
\end{center}
\vskip 0.1truein
\begin{center}
\bf{Pablo Carracedo,${}^1$\footnote{pablo.enrique.carracedo.garcia@xunta.gal}
Javier Mas,${}^2$\footnote{javier.mas@usc.es} Daniele Musso${}^2$\footnote{daniele.musso@usc.es} and 
Alexandre Serantes${}^2$\footnote{alexandre.serantes@usc.es}}\\
\end{center}
\vspace{0.5mm}

\begin{center}\it{
${}^1$meteo-Galicia, Santiago de Compostela, Spain}
\end{center}

\begin{center}\it{
${}^2$Departamento de  F\'\i sica de Part\'\i  culas \\
Universidade de Santiago de Compostela \\
and \\
Instituto Galego de F\'\i sica de Altas Enerx\'\i as (IGFAE)\\
E-15782 Santiago de Compostela, Spain}
\end{center}

\setcounter{footnote}{0}
\renewcommand{\thefootnote}{\arabic{footnote}}

\vspace{0.4in}

\begin{abstract}
\noindent
We construct a family of very simple stationary  solutions to gravity coupled to a massless scalar field in global AdS.
They involve a constantly  rising source for the scalar field at the boundary and thereby we name them pumping solutions. We construct them numerically in $D=4$. 
They are regular and, generically, have
negative mass. We perform a study of linear and nonlinear stability and find both stable and unstable branches. 
In the latter case, solutions belonging to different sub-branches can either decay to black holes or to limiting cycles.
This observation motivates the search for non-stationary exactly time-periodic solutions which we actually construct. 
We clarify the role of pumping solutions in the context of quasistatic adiabatic quenches. In $D=3$ the pumping solutions can be 
related to other previously known solutions, like magnetic or translationally-breaking backgrounds. From this we derive an analytic expression. 
\smallskip
\end{abstract}
\end{center}

\newpage

\tableofcontents


\section{Introduction}

Quenching a system is a way of probing its dynamics. 
Relevant information can be collected by studying the way the system responds and relaxes. 
This concerns time-dependent processes that in general have to be tackled with
numerical simulations.  Nevertheless, taking some specific limits allows to capture significant
features without very much computational cost. 
Recently, the limit of fast quenches has been under intense survey allowing to extract 
interesting scaling relations (see \cite{Das:2016eao} and references therein).
In the present paper we want to study the opposite limit, where considerable simplifications occur as well:
pumping solutions are very simple  configurations of the Einstein-scalar theory were the  real
massless scalar field has a constantly increasing source at the boundary. 

From the AdS/CFT perspective the pumping solutions are dual to an eternally monotonically driven conformal field theory. 
 Actually, although being strictly speaking a time dependent setup, the metric solution is perfectly static; technically 
 this is related to the fact that, in the massless  case, the metric couples to the scalar field only derivatively. 
 From the $d$-dimensional boundary theory viewpoint, the masslessness of the bulk scalar
 implies that the dual theory is monotonically deformed by  means of a marginal coupling. 

In this paper we will work  in the context of global AdS. For the planar situation, instead, one should first 
cure the infrared of the theory, and some considerations in the 
slow quench limit have been explored in \cite{Craps:2014eba} using the so called {\em hard wall} model. 
We find agreement with them in the existence of a maximum value for the driving time slope $\alpha_b$
and qualify this statement with some additional analysis.

 The present paper is organized in five sections. In Section 1 we  describe the stationary pumping ansatz and the equations to be solved.
We pause to clarify the difference of parameterizing the space of solutions in the origin-time or  the boundary-time gauge.
We describe the space of solutions and obtain their energy density.  This brings us to Section 2, where stability will be the main  concern, 
both at the linear and the nonlinear level. The former study is straightforward and neat, the latter is more articulate and brings in some interesting properties like
limiting cycles. Motivated by these oscillatory behaviors we search for and find time-periodic solutions, which we construct 
both perturbatively and numerically in Section 3. These periodic solutions branch continuously from  their analogs in the pure AdS geometry \cite{Maliborski:2013jca}.
In Section 4 we address the question of having a negative mass. We show that this effect is intimately tied to the fact that the scalar coupling rises linearly in time. 
Quenching back to a constant value for the scalar source restores the positivity of $m$. 
Section 5 focuses on the pumping solution over AdS$_3$ where, as usual, analytical control is much more attainable than in higher dimensions. In fact 
we provide a web of maps that links the pumping solution to a set of known  models in the literature. 
Notably this includes magnetic horizonless and regular as well as translationally-breaking solutions. 
This mapping proves to be extremely useful, allowing for an exchange of explicit results from one model to the other. 
In particular, in this way an analytical pumping solution is obtained. 
We close the paper with some supplementary material linking the new solution to those encountered in seemingly very different contexts. 
In Appendix A, we generalize the results of Section 5 to higher dimensions. At the level of the equations of motion, they
are shown to be derivable from a static magnetic $d$-form source.  Appendix B, intended for the specialized reader, provides technical material containing details of the numerical construction of time-periodic pumping solutions. 
 Finally in Appendix C we discuss a possible tension of the  pumping solution with respect to entanglement inequalities. We comment some caveats that need to be elucidated.

\section{Stationary pumping solutions}
\label{sps}

Let us consider a real massless scalar field coupled to gravity plus a negative cosmological constant in $d+1$ dimensions
\be
S = \int d^{d+1}x \sqrt{g}\left( \frac{1}{2\kappa^2} (R-2\Lambda)  - \frac{1}{2}\pd_\mu\phi \pd^\mu\phi\right)\, ,
\ee
with $\Lambda = -d(d-1)/2l^2$.
Hereafter we will set $l=1$, $\kappa^2 = d-1$ and fix the metric ansatz according to the following expression 
\be
ds^2 = \frac{1}{\cos^2\!x}\left( - f e^{-2\delta} dt^2+ f^{-1} dx^2 + \sin^2\!x \, d \Omega_{d-1}^2\right) \, , \label{line1}
\ee
where the radial coordinate spans $x\in [0,\pi/2)$ and all fields are, {\em a priori}, functions of $t$ and $x$.
The  equations of motion for the scalar field can be cast in terms of the fields $\Phi = \phi'$ and $\Pi = \dot\phi e^{\delta}/f$ as follows 
\be
\dot\Phi = \left(f e^{-\delta}  \Pi\right)'~~~;~~~\dot\Pi =\displaystyle \frac{1}{\tan^{d-1} x}\left(\tan^{d-1} x f e^{-\delta} \Phi\right)' \, .
\label{eoms}
\ee
From the Einstein equation we obtain just constraints
\beqa
\delta' &=&-\displaystyle \sin x \cos x \left(\Phi^2 + \Pi ^2\right) \, ,\label{eqd}\\
f' &=&\displaystyle \frac{d-2 + 2 \sin^2\!x}{\sin x \cos x} (1-f) +  \delta'  f \label{eqA} \, .
\rule{0mm}{8mm}
\label{constraints}
\eeqa
For $d\geq 3$, regularity of the Frobenius expansion close to the center of AdS$_{d+1}$  gives%
\footnote{In $d =2$ the field $f$ may have an arbitrary value $f_0\in {\mathbb R}$ at the origin.} 
\beqa
\phi(t,x) &=& \tilde\phi_0(t) +\tilde\phi_2(t) x^2 + ...\ , \nonumber\\
\delta (t, x) &=& \tilde\delta_0(t) + \tilde\delta_2(t) x^2  + ...\ ,  \label{norig} \\
f(t,x) &=& 1 +  \tilde f_2(t) x^2+ ...\ . \nonumber
\eeqa
Performing a similar expansion in $d=3$ for $x \to \pi/2$ and defining $\rho = \frac{\pi}{2}-x$, leads to the boundary series expansion\footnote{Note that for even $d$ the expansions contain in general logarithmic terms as well.}
\beqa
\phi(t,x) &=&\phi_{0}(t)+ \phi_{2}(t)\rho^2 + \phi_{3}(t) \rho^3 + ...~~ \nonumber\\
\delta(t,x) &=&\delta_0(t) +  \delta_{2}(t)\rho^2 + ...~~, \nonumber\\
f(t,x) &=& 1+f_{2}(t)\rho^2 + f_3(t) \rho^3 + ...  \label{expbound}
\eeqa
In this boundary expansion, every time-dependent coefficient is a functional of $\phi_0(t)$, $\phi_3(t)$, $\delta_0(t)$ and $f_3(t)$.  
In the AdS/CFT dictionary, $\phi_{0}(t)$ is dual to the QFT source, while $\phi_{3}(t)$ is proportional to the vev $\langle {\cal O}_\phi\rangle$.
Performing a coordinate change to bring the line element \eqref{line1} to the standard Fefferman-Graham form
$ds^2 =\frac{dz^2}{z^2} + \frac{g_{ab}(z,t,x)}{z^2} dx^a dx^b$  and expanding $g_{ab}(z,t,x) = \eta_{ab} + ... + z^3 t_{ab} + ...$ we can read off the value of the energy density $m\equiv \left<T_{tt}\right> = - 1/2 f_3$. 
Furthermore, the Ward identity associated to time translations reads\footnote{These statements rely on the holographic renormalization of the model which is commented briefly at the end of this section.}
\be
\partial_t \left<T_{tt}\right> = -\partial_t \phi_0 \langle {\cal O}_\phi \rangle \, . \label{mwi}
\ee

In order to look for background solutions, we plug the stationary ansatz $\dot\Pi=\dot\Phi=\dot f =\dot \delta =0$ 
 in \eqref{eoms}, and obtain the following constraints depending upon  two integration constants $\alpha$ and $\beta$
 \be
  \Pi = \frac{\alpha e^\delta}{f} \ , \qquad  \Phi =\frac{ \beta e^\delta}{f \tan^{d-1}\!x}\, . \label{eqsscal}
\ee
From the near origin expansion \eqref{norig} it directly follows that $\beta=0$, hence $\Phi=\phi'=0$ for all times. There exists a symmetry of \eqref{eqsscal} under $\delta \to \delta - a$ and $\alpha\to \alpha\, e^{a}$ that is the unbroken remnant, $t\to e^{-a} t$, of the group of time diffeomorphisms.  Tuning $a$ allows to select the position $x_p$ where $\delta(t,x_p)=0$.
We consider now the boundary time gauge choice, $x_p=\pi/2$, which is tantamount to setting $\delta_0(t) = 0$ in the near-boundary expansion \eqref{expbound}. 

We are interested in  solutions  characterized by a linearly rising source $\phi_0(t) = \alpha_b t$ where $t \in {\mathbb R}$. In the boundary time gauge, $t = t_b$,  these sourced solutions are the natural dual to a QFT observer performing continuous quenches. $\alpha_b$ is the pumping strength, and the label stresses the fact that its definition is linked to the boundary time gauge, where in addition one has $\Pi(t,\pi/2) = \alpha = \alpha_b$.

Inserting $\Pi$ and $\Phi=0$ into \eqref{eqd} and \eqref{eqA} we get a simple system to be solved for $f$ and $\delta$, namely
\beqa
\delta_b' &=&-\displaystyle \frac{ \alpha_b^2  \sin x \cos x e^{2\delta_b}}{f^2},  \label{eqds}\\
f' &=&\displaystyle \frac{d-2 + 2 \sin^2\! x}{\sin x \cos x} (1-f) +  \delta_b'  f \, . \label{eqAss}
\rule{0mm}{8mm}
\label{constraints}
\eeqa

In $d=2$  these equations admit an analytic solution, while for $d\geq 3$ they must be solved numerically. We will discuss the $d=3$ case from now on. We consider the boundary conditions $\delta(\pi/2)=0$, $f(\pi/2)=1$ and find that the equations can only be solved for $\alpha_b \leq \alpha_{b,max}$. In this sense, $\alpha_{b,max}$  corresponds to an intrinsic {\it adiabaticity threshold} present in our system. Here the concept
of  adiabaticity threshold refers to the fact that the source should grow slowly enough (with respect to the scale set by the $AdS$ radius, $\alpha_b^{-1}> \pi$) in order for the system to be able to keep up with it. Solutions with faster growth are either non-regular or non-static, typically collapsing into a black hole. In the following sections we further study this key point.

\begin{figure}
\center{\includegraphics[width=16cm]{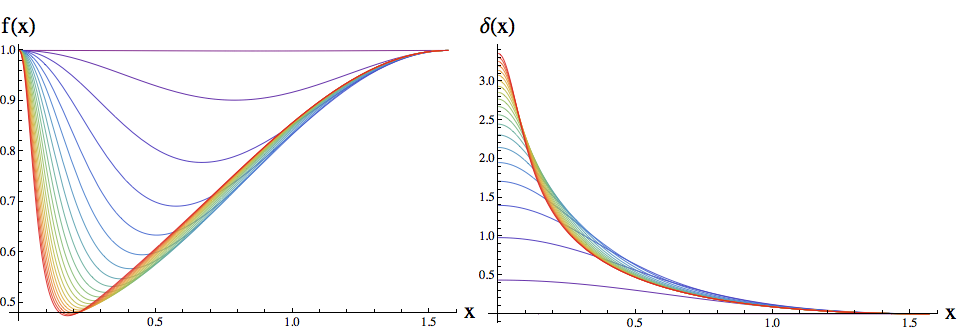}}
\caption{
\small Pumping solutions for $\alpha_o\in [0,20]$, the 
color frequency grows with growing $\alpha_o$.}
\label{fig:pumpingsols}
\end{figure}

\begin{figure}[h!]
\center{\includegraphics[width=16cm]{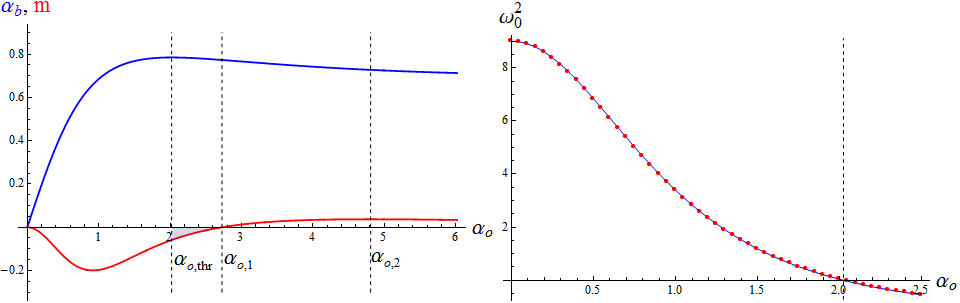}}
\caption{\small Left: $\alpha_b$ (blue) and energy density (red) of the pumping solution in terms of $\alpha_o$. The left dashed line corresponds to the maximum for $\alpha_b$ (to clarify the notation we recall that $\alpha_{b,\text{max}} = \alpha_b(\alpha_{o,\text{thr}})$ where $\alpha_b$ is the function plotted in blue). The central dashed line corresponds to $\alpha_o = \alpha_{o,1}$ below which the total mass is negative. The right dashed line corresponds to $\alpha_o = \alpha_{o,2}$ where the total mass attains its maximum value. The shaded region corresponds to the linearly unstable pumping solutions of negative energy density. Right: frequency of the fundamental mode of a linear radial fluctuation of the pumping solution. The linear instability threshold $\alpha_{o,\text{thr}}$ is signaled by the dashed vertical line.}
\label{fig:pumpingphase}
\end{figure}

Actually, the story is slightly subtler. It turns out that $\alpha_b$ is not a faithful parameter. 
Instead, it will be convenient to shift momentarily to the origin time gauge choice, where the time coordinate coincides with 
the proper time at the origin, $t=t_o$. 
The equations look exactly the same as \eqref{eqds} and \eqref{eqAss} replacing $\delta_b\to \delta_o$ and $\alpha_b\to \alpha_o$. The boundary condition to solve them now involves instead $\delta_o(x=0)=0$. 
It is clear that, in this gauge  $  \alpha_o = \Pi(x=0) $.
 Remarkably, it turns out  that, for {\em any} value of $\alpha_o\in [0, \infty)$ there exists a  stationary solution with vanishing $\langle {\cal O}_\phi \rangle$ and constantly rising scalar field $\phi(t) =\alpha_o t_o  = \alpha_b t_b $.
In Fig.\ref{fig:pumpingsols} we plot a family of solutions for regularly spaced values of $\alpha_0 \in [0,20]$.
The tension between the two parameterizations is resolved because the mapping $\alpha_o \to \alpha_b(\alpha_o)$ is not everywhere one-to-one. Actually, it has a  maximum at
$(\alpha_{o,{\rm thr}},\alpha_{b,{\rm max}}) = (2.01, 0.785)$.  This can be clearly seen in the left plot of Fig.\ref{fig:pumpingphase}. 

Let us pause here an discuss why the existence of the pumping solutions is remarkable. On general grounds, the pumping source $\phi_0(t)=\alpha t$ is expected to be associated with a process of energy injection that leads to a time-dependent geometry that eventually may collapse gravitationally. However, the Ward identity \eqref{mwi} readily implies that, if the pumping source is accompanied by a vanishing vev, $\left<\mathcal O_\phi(t)\right>=0$, the energy density remains constant, $\dot{m}(t) = 0$. This is precisely the situation in the pumping solution, since $\Phi(t,x)=0$. Notice that the physical mechanism behind the existence of the pumping solutions could also be the reason of its physical irrelevance: the question of whether the pumping solution background is infinitely fine-tuned, in the sense that it is destroyed by any infinitesimal perturbation such that $\Phi(t,x)\neq0$, needs to be explicitly addressed. 

In fact, we can compute the energy density of the pumping solutions, and perform two important observations (see Fig.\ref{fig:pumpingphase}, left). The first one is that there exist a threshold value $\alpha_{o,1}$ such that, for $\alpha_o < \alpha_{o,1}$, $m$ is {\it negative}. Similarly to other gravitational backgrounds with negative mass, this raises a concern regarding stability \cite{horowitz1998ads}. The second one is that there exists yet another threshold value $\alpha_{o,2} > \alpha_{o,1}$ where the energy density curve $m(\alpha_0)$ reaches an extremum. We may naively think that $\alpha_{o,2}$ implies the appearance of a Chandrasekhar instability. Both issues will be discussed in the next section. 

The identification of $f_3$ with the energy density $m$ relies as usual on a holographic renormalization program, including finite counter-terms. Although the full holographic renormalization of the time-dependent model is beyond the scope of the present analysis, we have checked that this identification is not affected in the present setup. Indeed, despite the breaking of time translations by the background, the counter-terms must be invariant combinations of the induced fields on the boundary cutoff surface. 
Their invariance is referred to the transverse diffeomorphisms. The only new counter-term which could affect the renormalized on-shell action linear in the fluctuations (\emph{i.e.} that could affect the definition of the energy density and the 1-point Ward identities) is proportional to
$\sqrt{-\gamma} (\partial_\mu \phi)(\partial^\mu \phi)$, where $\gamma$ is the induced metric on the boundary. Nevertheless, relying on the structure of the asymptotic near-boundary expansions of the fields, it is possible to show that such counter-term does not contribute at the finite level.%
\footnote{For a treatment of the holographic renormalization of massless scalars 
we refer in particular to \cite{Papadimitriou:2011qb}. Given the technical and conceptual connection of the present model with those breaking spatial translations, it is useful to relate the holographic renormalization in these two contexts, see \cite{Amoretti:2016bxs}.}

\section{Linear and nonlinear stability of the pumping solution}
\label{stab}

In order to connect two infinitesimally close pumping solutions through a normalizable fluctuation, one needs $\alpha_b'(\alpha_o) = 0$. Notice that this occurs at $\alpha_o=\alpha_{o,\text{thr}}$ and not at $\alpha_o=\alpha_{0,2}$ (where instead it is the mass which attains its maximal value). We recall that $\alpha_{b,\text{max}}$ corresponds to $\alpha_{o,\text{thr}}$.
For linearized fluctuations around the pumping solution contained in our original ansatz, an explicit computation shows that, indeed, at $\alpha_b = \alpha_{b,\textrm{max}}$, a zero mode appears in the eigenfrequency spectrum (see Fig.\ref{fig:pumpingphase}, right). Therefore, pumping solutions with $\alpha_o > \alpha_{o,\textrm{thr}}$ are linearly unstable. This entails that, in principle, only pumping solutions with $\alpha_o \leq \alpha_{o,\textrm{thr}}$ can be prepared by a quasistatic quench starting from the AdS$_4$ vacuum. 

Two questions naturally arise at this point. The first one is whether the linear stability of the $\alpha_o \leq \alpha_{o,\textrm{max}}$ pumping solutions translates into nonlinear stability. The second one concerns the final state reached by the $\alpha_o > \alpha_{o,\textrm{thr}}$ pumping solutions once perturbed. 

Let us address the first question. Notice first that, by virtue of the Ward identity \eqref{mwi}, even if a linear eigenmode has a nonzero $\left<\mathcal O_\phi(t) \right>$, since its time dependence is harmonic, the energy density it generates would also oscillate periodically with a $\pi/2$ phase shift with respect to $\left<\mathcal O_\phi(t) \right>$. Nonlinear perturbations, being generically nonharmonic, might trigger a nonlinear instability due to the existence of a finite $\Phi(t,x)$ with generic time dependence, that can lead to an energy exchange eventually resulting in a gravitational collapse. There exist two complementary ways of establishing the absence of potential nonlinear instability. The first one is realizing that pumping solutions below $\alpha_{o,\textrm{thr}}$ can be actually accessed through a quasistatic quench. Indeed, we have simulated the time-dependent geometry generated by a source $\phi_0(t)$ such that $\phi_0(t<0)=0$, $\ddot{\phi_0}/\dot{\phi_0}^2 \ll 1$ and $\lim_{t\to\infty} \phi_0/t = \alpha_b$, and have checked that, at sufficiently late times, it settles down into the linearly stable pumping solution corresponding to $\alpha_b$ (see Fig.\ref{fig:adiabquench} in Section 5). In consequence, the pumping solutions play an important role at the nonlinear level: were they subjected to nonlinear instability, the system would never relax to them.
\begin{figure}[ht]
\begin{center}\includegraphics[width=16cm]{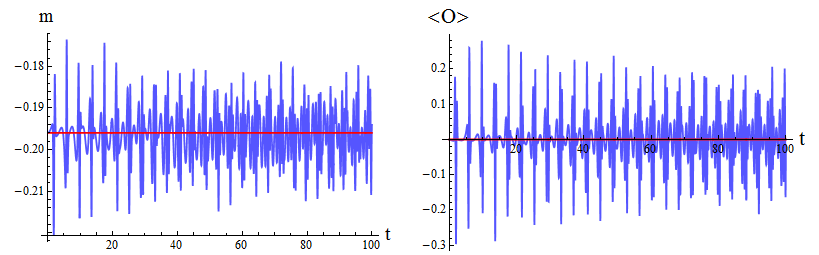}
\end{center}
\caption{\small Energy density (left, blue) and vev (right, blue) for the $\epsilon=1, \sigma=0.1$ Gaussian perturbation of a pumping solution with $\alpha_o=1$. Red curves denote the initial values of each plotted quantity. In this simulation, the absolute error in the Ward identity is bounded by $10^{-4}$ for a grid resolution $N=2^{11}$.}
\label{fig:gausspumping}
\end{figure}

The second approach we followed in order to determine whether linearly stable pumping solutions are also nonlinearly stable was to take, as initial data at $t =0$, 
\beqa
&&\Phi(0,x)=0, \\
&&\Pi(0,x) = \Pi_{\alpha_o}(x) + \frac{2\epsilon}{\pi} \exp\left( -\frac{4 \tan^2 x}{\pi^2\sigma^2} \right)\, , \label{gausspumping}
\eeqa
and simulate their subsequent time evolution. $\Pi_{\alpha_o}(x)$ represents a pumping solution characterized by $\alpha_o$. An example is depicted in Fig.\ref{fig:gausspumping}, for $\alpha_o=1$, $\epsilon = 1$ and $\sigma=0.1$. Due to the Gaussian perturbation, the initially negative $m(0)$ does not remain constant: its time derivative perfectly satisfies the Ward identity \eqref{mwi}  in the presence of the nontrivial $\left<\mathcal O_\phi\right>$ that is generated. Both quantities, $m(t)$ and $\left<\mathcal O_\phi(t)\right>$, are found to oscillate in a disordered way around their respective initial values. In fact, we find that the time-averaged energy density, $\left<m\right> = t^{-1} \int_0^t m(t') dt'$, rapidly reaches a constant value and the system absorbs/loses no net energy density. Therefore, no sign of nonlinear instability is found.

\subsection{The linearly unstable branch}

This situation changes dramatically for the linearly unstable branch of pumping solutions. In fact, repeating the procedure described above for finite $\epsilon$, a black hole promptly forms. Instead, initializing the simulation code with $\epsilon = 0$, puts us exactly on top of a $\alpha_o > \alpha_{o,\textrm{thr}}$ pumping solution, up to numerical error. Being linearly unstable, this unavoidable numerical error drives the system away from the original geometry. However, for $\alpha_{o,\text{thr}}<\alpha_o\leq \alpha_{o,c}$, despite entering a time-dependent regime, the system does not undergo gravitational collapse. It rather decays into a limiting cycle that, for $0<\alpha_o - \alpha_{o,\text{thr}} \ll 1$, is manifestly time-periodic.\footnote{The quantity $\alpha_{0,c}$ is close to, but distinct from, the value $\alpha_{o,2}$ that marked the sign change of the energy density of the pumping solution.} 

\begin{figure}[h]
\center{\includegraphics[width=16cm]{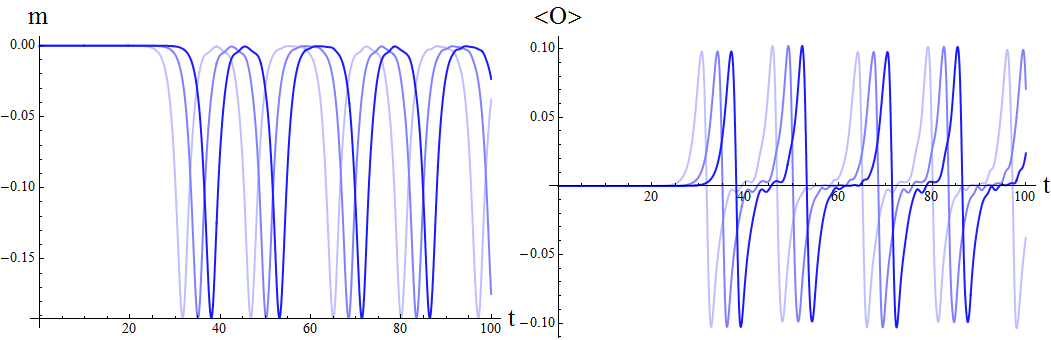}}
\caption{\small Time evolution of the mass (left) and vev (right) for the oscillatory solution with $\alpha_o = 2.71$. The coloring encodes the three spatial resolutions $N=2^{10}$, $2^{11}$, $2^{12}$ employed to build the initial data and perform the simulation. Consistently, the transition to a time-dependent solution happens at later times for increasing $N$. On the contrary, the periodicity is resolution-independent. This means that the system reaches the same limiting cycle at different times depending on the initial conditions.}
\label{unsta}
\end{figure}

 Let us perform two different checks that help establishing that the limiting cycle solutions are not  a numerical artifact. First, since the departure from the original unstable pumping solution is noise-driven, the transient time needed to reach the limiting cycle solution should be proportional to the grid resolution. We find that this is precisely what happens (see Fig.\ref{unsta}). 

\begin{figure}[h!]
\center{\includegraphics[width=16cm]{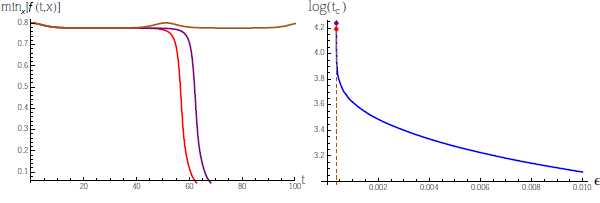}}
\caption{\small Left: $\textrm{min}_x f(t,x)$ for the perturbed limit cycle solution for three deformations of the form \eqref{limit_cycle_def} close to $\epsilon_c$, two above (red and purple) and one below (brown). Right: $t_c(\epsilon)$ for the perturbed limit cycle solution (blue). The purple and red dots correspond to the purple and red simulations depicted on the left figure. The brown dashed line corresponds to the brown simulation on the left, for which no collapse is found.}
\label{limit_cycle_stability}
\end{figure}

As a second check, we wait until the system has decayed into the limiting cycle solution, and take the $\Phi(t_{ref}, x), \Pi(t_{ref}, x)$ profiles at the reference time $t_{ref}$ as initial data of another numerical simulation where the deformation 
\beq
\Phi(t_{ref}, x) \rightarrow \Phi(t_{ref},x) + \epsilon \cos^2 x \sin x \label{limit_cycle_def}
\eeq
is implemented. Then, we perform a scan in $\epsilon$, with the aim of seeing if the perturbed limiting cycle solution collapses gravitationally. The results of this analysis for the unstable pumping solution with $\alpha_b = 0.78497$ can be found in Fig.\ref{limit_cycle_stability}. The collapse time $t_c(\epsilon)$ looks divergent at a critical $\epsilon_c$ and, for $\epsilon \leq \epsilon_c$, no black hole forms within the times computationally accessed. 

\section{Time-periodic pumping  solutions}\label{sec_time_periodic}

The endpoint of the unstable pumping branch for $\alpha_o<\alpha_{o,c}$ is, as said before, a limiting cycle. For $0 < \alpha_o -\alpha_{o,{\rm thr}} \ll 1$ this limiting cycle has a single period. This fact came as unexpected, and prompted us to look for exactly periodic solutions with a pumping source. Time-periodic solutions in global AdS supported by a real massless scalar field with zero source were originally presented in \cite{Maliborski:2013jca} and we now show that, for $\alpha_b \neq 0$,\footnote{For  small $\alpha_{o}$ we can  equally use $\alpha_b$ as deformation parameter since the relation that links them is one-to-one.} each such solution generalizes into a family of exactly periodic pumping solutions. The existence of pumping time-periodic solutions can be first proved at the perturbative level and later corroborated with a numerical approach.

Let $\Omega$ be the frequency of the periodic solution we aim at.\footnote{Recall that scalar eigenmodes in global AdS have a 
frequency spectrum given by $\omega_n = \Delta + 2 n$, where $\Delta$ is the conformal dimension 
of the dual scalar operator. For a massless scalar field in AdS$_4$, $\Delta = 3$.} Rescale the time coordinate as $\tau \equiv \Omega\, t$ and introduce the following ansatz
\beqa
&&\phi(\tau, x) = \alpha_b \frac{\tau}{\Omega(\alpha_b, \epsilon)} + \epsilon \cos^3\!x \sin \tau + \sum_{n = 1}^\infty \sum_{k=0}^{2n+1}\varphi_{2n+1-k,k}(\tau, x) \alpha_b^{2n+1 - k} \epsilon^k, \label{expansion} \label{periodic_perturbative_phi}\\
&&\delta(\tau, x) = \sum_{n = 1}^\infty \sum_{k=0}^{2n} d_{2n - k, k}(\tau, x) \alpha_b^{2n-k}\epsilon^k, ~f(\tau, x) = 1 + \sum_{n = 1}^\infty \sum_{k=0}^{2n} a_{2n-k, k}(\tau,x) \alpha_b^{2n-k}\epsilon^k, \label{periodic_perturbative_metric}\\
&&\Omega(\alpha_b, \epsilon) = 3 + \sum_{n = 1}^\infty \sum_{k=0}^{2n} \omega_{2n-k}  \alpha_b^{2n-k}\epsilon^k. \label{periodic_perturbative_frequency}
\eeqa
At first order in $\alpha_b$ and $\epsilon$ we are just considering the linear superposition of the pumping solution $\phi_\alpha(t,x) = \alpha_b t$ and the fundamental eigenmode of the scalar field $\phi_\epsilon = \epsilon \sin \tau \cos^3\!x$ over global AdS$_4$. Higher order corrections determine how the seed $\phi_\alpha + \phi_\epsilon$ is dressed nonlinearly. 
The interested reader can find a thorough discussion with technical details in Appendix \ref{pumping_periodic}. For our present purposes it suffices to mention that, after the first nonlinear correction, the seed frequency  $\Omega_0 = 3$ is modified to
\beq
\Omega = \Omega_0 - \frac{135}{128}\epsilon^2 - \frac{7}{4}\alpha_b^2. \label{Omega_perturbative} 
\eeq
To close the gap between the perturbative and the fully nonlinear regime, one needs to resort to numerical methods. 
In particular, since the eigenmodes of the AdS$_4$ Laplacian do not satisfy the boundary conditions \eqref{expbound},\footnote{A finite mass breaks the parity under $\rho \to - \rho$ that the scalar field near-boundary expansion exhibits at the linearized level.} we have adapted the pseudospectral method described in \cite{Maliborski:2016zlh} to our present setup. 
 
\begin{figure}[h!]
\begin{center}
\includegraphics[width = 16cm]{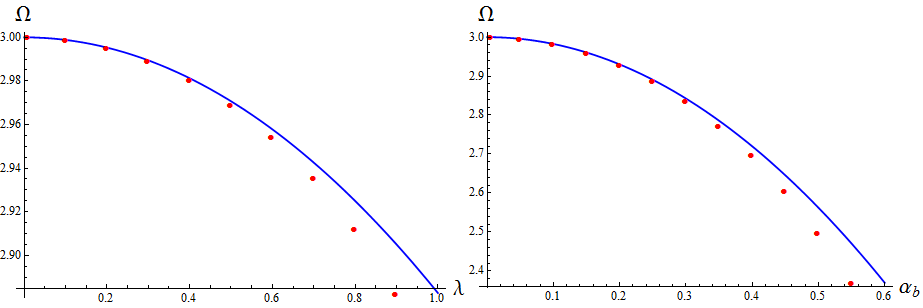}
\end{center}
\caption{\small {Left: $\Omega(0.0005, \lambda)$ as determined numerically (red dots) and perturbatively (blue line). Right: $\Omega(\alpha_b, 0.001)$ as determined numerically (red dots) and perturbatively (blue line).}} 
\label{fig:oscillon-frequency}
\end{figure}
 
Let us perform two different checks on the periodic solutions found numerically. We define $\lambda$ as the value of the vev at $\tau =\pi/2$,  $\lambda \equiv \left<\mathcal O_\phi(\tau=\pi/2) \right> = 3 \epsilon$. In Fig.\ref{fig:oscillon-frequency}, we plot $\Omega(\alpha_b, \lambda)$ for two orthogonal directions in the phase space of exactly periodic solutions, and compare it with \eqref{Omega_perturbative}. In each case, excellent agreement is found between the perturbative and the numerical result in the $\alpha_b, \epsilon \ll 1$ regime. 

\begin{figure}[h!]
\begin{center}
\includegraphics[width = 16cm]{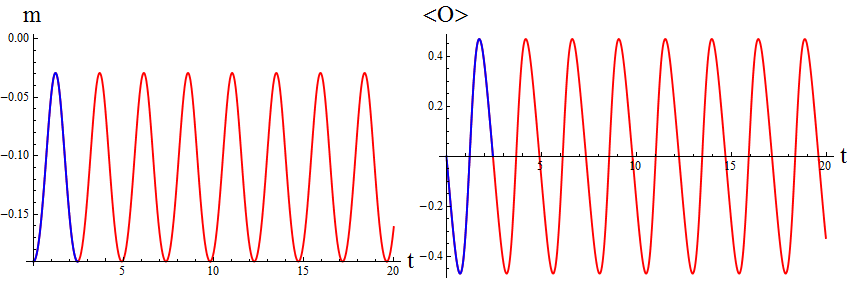}
\end{center}
\caption{\small {Energy density and vev for an exactly periodic solution with $\alpha_b = 0.45$, $\lambda = -0.45$ 
as obtained from the pseudospectral generating code (blue) and the simulated time evolution of the initial data (red).}}  
\label{fig:oscillon-evolution}
\end{figure}

As a further consistency check, we perform a simulation starting from $\Pi(t=0,x) = \Pi_0(x)$ and $\Phi(t=0,x) = 0$ 
as determined by the pseudospectral algorithm and study their subsequent time evolution. We provide one example in Fig.\ref{fig:oscillon-evolution}. It is observed that the time development of the initial data perfectly agrees with the output of the pseudospectral code, providing a highly nontrivial check of our numerical procedures. 

\section{Slow and fast quenches}

We have already mentioned that all pumping states with $\alpha_b\leq \alpha_{b,{\rm max}}$ can be prepared by a 
boundary quasistatic quench
 $\alpha_{b}(t)$. Notice that the  stability analysis in Section \ref{stab} was restricted to isotropic perturbations.  We have seen that the fact that the pumping solution has energy density below that of AdS is not associated to an instability for small values of $\alpha_o$.
\begin{figure}[ht]
\begin{center}
\includegraphics[scale=0.48]{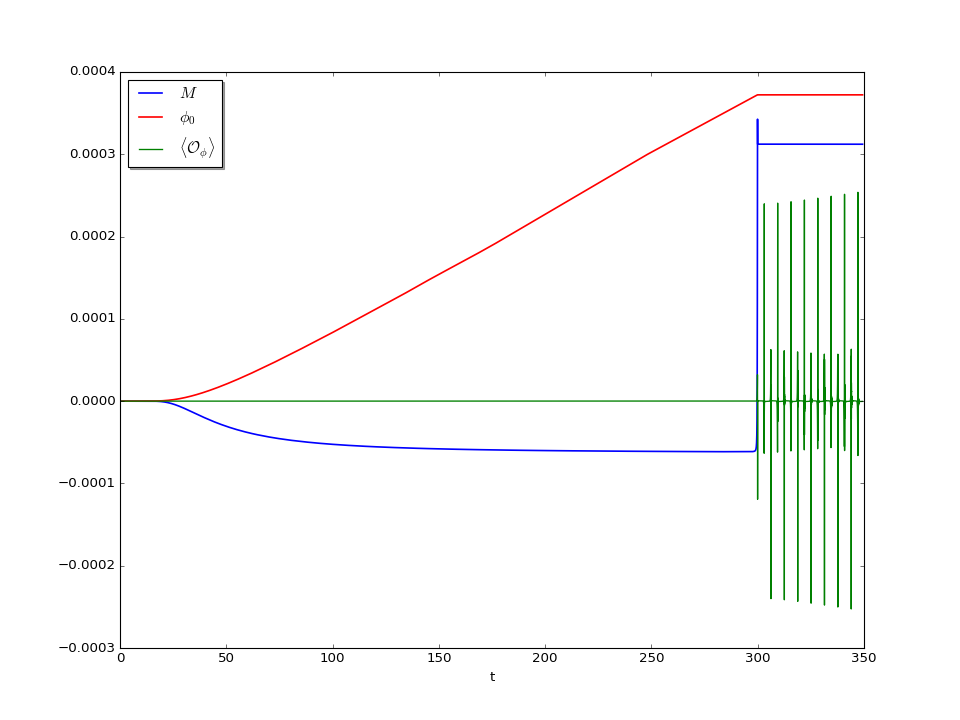}
\end{center}
\caption{\small Quenching off the pumping solution. The time span $\delta$ for the interpolation has been set to  $\delta_i=30$ for a quasi-static build up of the pumping solution and $\delta_f = 0.1$ for the switching off. The actual values for $\phi_0$ and $\langle\mathcal O_\phi\rangle$ have been rescaled to fit into the same figure. From the vev $\langle {\cal O}_\phi\rangle$ it is apparent that the final state is dynamical.}
\label{fig:adiabquench}
\end{figure}
\\
A nice consequence of this is the fact that the pumping solution with negative mass can be obtained by means of a  quench that starts from the AdS vacuum, $\alpha_b=0$, and ramps this slope
quasistatically to some  $0<\alpha_b<\alpha_{b,max}$. This can be observed in the first part of Fig.\ref{fig:adiabquench}, where the pumping solution is the mid part ($100<t<300$). This  experimental way of building the pumping solution illustrates also the meaning of the adiabadicity threshold.  Indeed, ramping $\alpha_b$ beyond the value $\alpha_{b,max} = 0.783$ leads immediately to sudden black hole formation.
In other words, there is no way to explore the region beyond $\alpha_{b,max}$ by means of an experimental quench. We also want to remind the reader that all the solutions that in Fig.\ref{fig:pumpingphase} lie to the right of the maximum of the blue curve are
eternal and only exist as solutions to the static equations. \\ 
Turning back to the issue of the possible puzzle with having a stable static solution with negative mass,
a  numerical experiment is at the same time clarifying and intriguing.  
Starting from a pumping solution with negative mass, like the one in Fig.\ref{fig:adiabquench} at intermediate times ($100<t<300$), we can think of  suddenly isolating the system from the environment  by switching off the scalar source. Naively, from the Ward identity, no energy inflow can occur and one would think that, in this way, we end up with  a a regular stationary solution of negative mass. However, this does not happen. Turning off the pumping can be achieved by stopping the linear growth of $\phi_0$ as some time $t_f$.  In practice, this has to be implemented through a smooth interpolation of time span $\delta$ between the pumping boundary condition $\dot\phi_0(t < t_f-\delta/2) = \alpha_b $ and a constant final value $\dot \phi_0(t >t_f+\delta/2 ) = 0$ in the limit $\delta \to 0$. The remarkable result is that, for any value of $\delta$ and any interpolation used, the solution always returns to a positive energy density $m>0$ at times $t>t_f+\delta$. In fact, for an extremely wide and smooth interpolation, the final configuration attained is the static AdS$_4$ background with $m=0$. 

For more abrupt interpolations the eventual geometry is time-dependent with a 
constant positive mass $m>0$. An example for $\delta = 0.1$ can be seen in Fig.\ref{fig:adiabquench}. At late times, the final fate of this evolution encounters the issue of stability of a scalar pulse evolving in global AdS$_4$. In fact, if we consider the limit $\delta\to0$, the energy density peaks so violently that a direct collapse may occur immediately after $t_f$. 

\begin{figure}[h]
\begin{center}
\includegraphics[width=16cm]{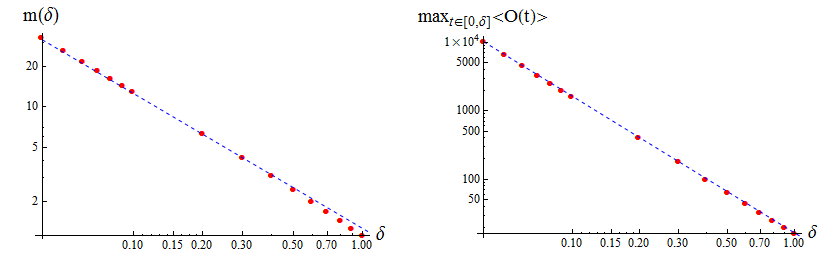}
\end{center}
\caption{\small Left: log-log plot of $m(\delta)$ (red dots) and the fitting curve $m(\delta) \propto \delta^{-1}$ (blue dashed) for a quench off the $\alpha_0=1$ pumping solution with interpolating function \eqref{int_funct}. Right: log-log plot of $\textrm{max}_{t\in[0,\delta]}\left<\mathcal O_\phi(t)\right>$ (red dots) and the fitting curve $ \textrm{max}_{t\in[0,\delta]}\left<\mathcal O_\phi(t)\right>\propto \delta^{-2}$ (blue dashed) for the same quench process as in the left plot.}
\label{fig:quenching_off}
\end{figure}

Let us analyze the above mentioned behavior in detail. In Fig.\ref{fig:quenching_off}, we plot $m(\delta)$ and $\textrm{max}_{t\in[0,\delta]}\left<\mathcal O_\phi(t)\right>$ for a quench off the $\alpha_o = 1$ pumping solution with the interpolating function 
\beq
\dot{\phi}_0(t) = \frac{1}{2}\alpha_b\left[1+\tanh\left(\frac{\delta}{t}+\frac{\delta}{t-\delta}\right)\right]\qquad , \qquad  t\in[0,\delta].\label{int_funct}
\eeq
For $\delta \lesssim 1$, both $m(\delta)$ and $\textrm{max}_{t\in[0,\delta]}\left<\mathcal O_\phi(t)\right>$ display a well defined scaling with respect to $\delta$, $m(\delta) \propto \delta^{-1}$ and $ \textrm{max}_{t\in[0,\delta]}\left<\mathcal O_\phi(t)\right> \propto \delta^{-2}$, confirming that both diverge in the $\delta \to 0$ limit. 

\begin{figure}[h]
\begin{center}
\includegraphics[width=16cm]{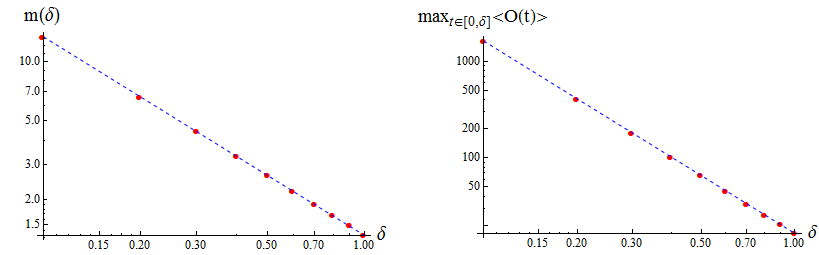}
\end{center}
\caption{\small Left: log-log plot of $m(\delta)$ (red dots) and the fitting curve $m(\delta) \propto \delta^{-1}$ (blue dashed) for the quench profile \eqref{int_funct_2} with $\alpha_b$ as in \eqref{int_funct}. Right: log-log plot of $\textrm{max}_{t\in[0,\delta]}\left<\mathcal O_\phi(t)\right>$ (red dots) and the fitting curve $ \textrm{max}_{t\in[0,\delta]}\left<\mathcal O_\phi(t)\right>\propto \delta^{-2}$ (blue dashed) for the same quench process as in the left plot.}
\label{fig:quenching_on}
\end{figure}

A natural question is if the observed scaling is related to the fact that we are quenching off the pumping solution, or is rather determined by the particular form of the interpolating function \eqref{int_funct}. To answer this question, in Fig.\ref{fig:quenching_on} we plot $m(\delta)$ and $\textrm{max}_{t\in[0,\delta]}\left<\mathcal O_\phi(t)\right>$ for a quench with $\phi_0(t)$ given by the time-reversed version of \eqref{int_funct} acting over the AdS$_4$ vacuum, 
\beq
\dot{\phi}_0(t) = \frac{1}{2}\alpha_b\left[1-\tanh\left(\frac{\delta}{t}+\frac{\delta}{t-\delta}\right)\right] \qquad , \qquad t\in[0,\delta].\label{int_funct_2}
\eeq
We observe that the former scaling relations $m(\delta) \propto \delta^{-1}$, $ \textrm{max}_{t\in[0,\delta]}\left<\mathcal O_\phi(t)\right> \propto \delta^{-2}$ uphold, showing that the scaling is solely determined by the precise non-analytic behavior of the source in the $\delta \to 0$ limit: both \eqref{int_funct} and \eqref{int_funct_2} result in quench profiles $\phi_0(t)$ that are continuous but non-differentiable.  

\begin{figure}[h]
\begin{center}
\includegraphics[width=16cm]{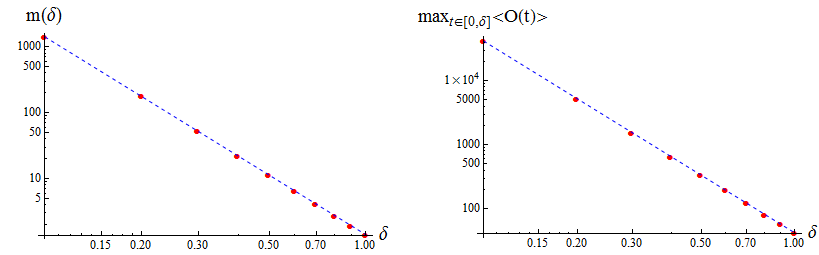}
\end{center}
\caption{\small Left: log-log plot of $m(\delta)$ (red dots) and the fitting curve $m(\delta) \propto \delta^{-3}$ (blue dashed) for a quench profile given by the {\it rhs} of \eqref{int_funct_2} applied to $\phi_0(t)$. Right: log-log plot of $\textrm{max}_{t\in[0,\delta]}\left<\mathcal O_\phi(t)\right>$ (red dots) and the fitting curve $ \textrm{max}_{t\in[0,\delta]}\left<\mathcal O_\phi(t)\right>\propto \delta^{-3}$ (blue dashed) for the same quench process as in the left plot.}
\label{fig:buchel}
\end{figure}

The scaling displayed by $m(\delta)$ and $\textrm{max}_{t\in[0,\delta]}\left<\mathcal O_\phi(t)\right>$ under \eqref{int_funct_2} differs from the one that would be present if the quenching profile \eqref{int_funct_2} were considered  on $\phi_0(t)$, rather than on  $\dot{\phi}_0(t)$. In the case of a $\phi_0(t)$ fast quench, the scaling relations are $m(\delta) \propto \delta^{-3}$ and $ \textrm{max}_{t\in[0,\delta]}\left<\mathcal O_\phi(t)\right> \propto \delta^{-3}$, as expected on purely dimensional grounds \cite{Buchel:2013gba}. We plot representative results of this behavior in Fig.\ref{fig:buchel}.  

These numerical experiments confirm that no negative mass will be measured by a boundary observer when the source has settled to a constant value. It rises therefore the question about the nature of the quantum state dual to the pumping geometry. Is it really a {\em bona fide} quantum state? We will see in Appendix C that further information coming from entropy inequalities may cast some further shadows on this. 

As a further observation, note that the whole process of quenching in and off embodies some irreversible character.
Imagine slowly building up the value of $\alpha_b(t)$ from zero to a constant final value in a time span $\delta$,  staying there for a long time,  and finally switching back $\alpha_b(t) \to 0$ using the same $\delta$-smeared step function, but time reversed. Even if both the equations of motion and the boundary conditions are time reversal invariant, the end result will always have $m(t=\infty)\geq m(t=0)=0$, and only $m(t=\infty)=0$ in the quasistatic limit $\delta \to \infty$ for both steps.

\section{Pumping solution over AdS$_3$}
\label{commu}

In this section we shall investigate the three dimensional case. It turns out that the pumping solution over AdS$_3$ is related to a number of other solutions known in the literature via a series of mappings that includes double Wick rotations
and/or Hodge dualities. This observation allows us to give an analytic expression for the pumping solution. 

Let us consider a three-dimensional charged black brane with negative cosmological constant \cite{Banados:1992wn,Banados:1992gq}. The metric and the electromagnetic potential are given by 
\beqa
ds^2 &=& - h(r) dt^2 + \frac{dr^2}{h(r)}+r^2 dx^2, \label{ds2_RN} \\
h(r) &=& r^2 - M - Q_e^2 \log(r),\label{f_RN} \\
\rule{0mm}{6mm}
A(r) &=& \left[\mu - Q_e \log(r)\right]dt. \label{A_RN}
\eeqa
In the expressions above, $r \in [0, \infty)$ and $t, x \in \mathbb R$. $M$ and $Q_e$ determine respectively the mass and the electric charge of the black brane geometry.  From now on, we focus on the one-parameter family of charged black branes with $M = 1$ and $\mu = 0$. Note that the outer horizon is located at $r = r_h = 1$ for $\left|Q_e\right| < \sqrt{2}$. The temperature is
\beq
T = \frac{1}{2\pi} \left(1-\frac{1}{2}Q_e^2\right). \label{temperature}
\eeq
At $\left|Q_e\right| = \sqrt{2}$ the black brane is extremal. For $\left|Q_e\right| > \sqrt{2}$, the hypersurface $r = 1$ actually corresponds to the inner horizon, and \eqref{temperature} does not represent the real temperature of the back brane. By performing the coordinate change $r^2 = 1 + R^2$, the exterior region of the charged black brane can be parameterized as 
\beqa
ds^2 &=& - h(R) dt^2 + \frac{R^2 }{(1+R^2)h(R)}dR^2+ (1+R^2) dx^2,\label{ds2_charged_ext}\\
\rule{0mm}{6mm}
h(R) &=& R^2  - \frac{1}{2}Q_e^2 \log(1+R^2), \\
\rule{0mm}{7mm}
A &=& -\frac{1}{2}Q_e \log(1+R^2)dt. 
\eeqa
with $R \in [0, \infty)$. Now, let us perform the double Wick rotation defined by 
\beq
t \to i \varphi,~~~x \to i t,~~~~Q_e \to  -i Q_m, \label{double_Wick}
\eeq
in a such a way that the geometry and electromagnetic potential of the exterior region are mapped to the following solution of three-dimensional Einstein-Maxwell theory with a negative cosmological constant
\beqa \label{geo1}
ds^2 &=& - (1+R^2) dt^2 + \frac{R^2 }{(1+R^2)h(R)}dR^2+ h(R) d\varphi^2,\\ \label{geo2}
h(R) &=& R^2  +\frac{1}{2}Q_m^2 \log(1+R^2), \label{f_magnetic_0}\\
\rule{0mm}{7mm}
A &=& -\frac{1}{2}Q_m\log(1+R^2)d\varphi.\label{A_magnetic_0}
\eeqa
This is a horizonless geometry. It can be explicitly shown that the coordinate $\varphi$ must be identified with period 
$\beta = 2\pi/(1+Q_m^2/2)$ in order to avoid a conical singularity at $R = 0$. On the other hand, this period 
is nothing but the analytic continuation of the inverse temperature of the original charged black brane under the double 
Wick rotation. The geometry \eqref{geo1}, \eqref{geo2} is supported by the radial magnetic field associated to \eqref{A_magnetic_0}, and we will refer to it as the {\it magnetic} AdS$_3$ solution.\footnote{This background was originally found in references \cite{Clement:1993kc,hirschmann1996magnetic,Cataldo:1996yr}. In \cite{Cataldo:1996yr}, the authors introduced the double Wick rotation we discussed here. See \cite{dias2002rotating,olea2005charged} for the construction of the general magnetic solution with angular momentum, the computation of its conserved charges and a proposed physical interpretation. The physical nature of the magnetic solution was further elucidated in \cite{cataldo2004magnetic}.} 

We want to map the magnetic solution to the pumping solution. To this purpose we will rely on Hodge duality. However we have still to massage the magnetic solution: let us rescale our coordinates as $t \to \beta/(2\pi)t$, $R \to 2\pi/\beta R$ 
and $\varphi \to \beta/(2\pi)\varphi$, in such a way that  $\varphi \sim \varphi + 2 \pi$.%
\footnote{The rescaling of the $t$ and $R$ coordinates follows from the requirement that $\lim_{R\to\infty} R^{-2}ds^2 = - dt^2 + d\varphi^2$.
This corresponds to the standard flat space metric as the representative of the conformal structure of the boundary where we have chosen $\Omega(R) = R^{-2}$ as conformal factor. Note also that this is the same procedure we would have employed were we on the vacuum state.} In this new coordinate system, we have that 
\beqa
ds^2 &=& - \left[\left(\frac{\beta}{2\pi}\right)^2+R^2\right] dt^2 + \frac{R^2 }{\left[\left(\frac{\beta}{2\pi}\right)^2 +R^2\right]h(R)}dR^2+ h(R) d\varphi^2,\\
\rule{0mm}{8mm}
h(R) &=& R^2 + \frac{1}{2}\tilde{Q}_m^2 \log\left[\left(\frac{\beta}{2\pi}\right)^2+R^2\right]+\tilde{\mu}\tilde{Q}_m,\label{f_magnetic}\\
\rule{0mm}{8mm}
A &=& -\left\{\tilde{\mu} + \frac{1}{2}\tilde{Q}_m\log\left[\left(\frac{\beta}{2\pi}\right)^2+R^2\right]\right\} d\varphi,  \label{A_magnetic}
\eeqa
where we have defined
\beq
\tilde{Q}_m \equiv \frac{\beta}{2\pi}Q_m = \frac{Q_m}{1+\frac{1}{2}Q_m^2} \, \qquad ,\qquad
\tilde{\mu} \equiv \tilde{Q}_m \log \frac{2\pi}{\beta} \ .
\eeq
The field strength corresponding to the magnetic solution is 
\beq
F = -\tilde{Q}_m\frac{R}{\left(\frac{\beta}{2\pi}\right)^2+R^2}\; dR \wedge d\varphi. \label{F_magnetic}
\eeq

Under Hodge duality, the magnetic AdS$_3$ solution can be put into correspondence with the three-dimensional pumping solution. Recall that the vacuum Maxwell equations are $dF = d(\star F) = 0$, where $\star$ is the Hodge operator. 
In three dimensions, the Hodge dual of a two-form corresponds to a one-form. Hence $F$ can 
be alternatively expressed as $F = \star d\phi$, where $\phi$ is a massless scalar field. The Bianchi identity 
$dF=0$ transforms into the scalar field equation of motion, $-\star d(\star d\phi) = \nabla^2 \phi = 0$, 
while the dynamical Maxwell equation reduces to the trivial statement $d^2 \phi = 0$. 
The map leaves the metric invariant and, as a consequence, a metric supported by a given electromagnetic field can be thought of as being sourced by the dual scalar field defined by this procedure.  

Let us consider explicitly how the Hodge duality works at the level of the solution. Take the three-dimensional pumping solution with $\phi(t, r) = \alpha t$. Under Hodge duality, we find that the scalar field profile maps to
\beq
F =  - \alpha  \frac{ r e^{\delta(r)}}{f(r)} dr \wedge d\varphi, \label{pumping_field_strength}
\eeq
where we have parameterized the pumping solution geometry in standard Schwarzschild coordinates, 
\beq \label{shwa}
ds^2 = - f(r) e^{- 2 \delta(r)} dt^2 + \frac{dr^2}{f(r)} + r^2 d\varphi^2. 
\eeq
Note that the radial coordinates of the magnetic AdS$_3$ solution and the pumping solution are related by
\beq
r^2 = h(R), \label{r_R}
\eeq
as it emerges from comparing the $\varphi\varphi$-component of the metric \eqref{geo1} and \eqref{shwa}.
From a comparison of the radial components of \eqref{geo1} and \eqref{shwa}, we have that $r$ and $R$ must also satisfy
\beq
\frac{dr^2}{f(r)} =\frac{R^2 dR^2}{\left[\left(\frac{\beta}{2\pi}\right)^2+R^2\right]h(R)}. 
\eeq

Deriving \eqref{r_R} and substituting into the expression above, we obtain the relation 
\beq
f(r(R)) =  \left[\left(\frac{\beta}{2\pi}\right)^2+R^2\right]\left(1+ \frac{1}{2}\frac{\tilde{Q}_m^2}{\left(\frac{\beta}{2\pi}\right)^2+R^2}\right)^2\ , \label{h_R}
\eeq
where the explicit form of $h(R)$ provided by equation \eqref{f_magnetic} has been employed. Note that, when $R = 0$, \eqref{f_magnetic} implies that $r = h(0) = 0$. As a consequence, from \eqref{h_R} and the definitions of $\beta$ and $\tilde{Q}_m$, it follows that $f(0) = 1$, a fact that shows the absence of a conical singularity at the origin in a manifest way. 

By identifying now the $tt$-components of the metrics \eqref{geo1} and \eqref{shwa}, we get that 
\beq
\left(\frac{\beta}{2\pi}\right)^2+R^2 = f(r) e^{-2 \delta(r)}. 
\eeq
Employing equation \eqref{h_R}, we finally obtain
\beq
\delta(r(R)) = \log\left(1 + \frac{1}{2}\frac{\tilde{Q}^2_m}{\left(\frac{\beta}{2\pi}\right)^2+R^2}\right). \label{d_R}
\eeq
Notice that $\delta(\infty) = 0$, which means that the parametrization of the magnetic solution corresponds to the boundary time gauge (see Section \ref{sps}). 

With the help of equations \eqref{r_R}, \eqref{h_R} and \eqref{d_R} we can express the field strength dual (in the Hodge sense) to the pumping solution in $R$ coordinates, namely
\beq 
F = -\alpha\frac{R}{\left(\frac{\beta}{2\pi}\right)^2+R^2}\; dR \wedge d\varphi. \label{F_pumping}
\eeq
Comparing \eqref{F_pumping} with \eqref{F_magnetic} we find that the pumping solution corresponds to a magnetic solution such that
\beq
\alpha = \tilde{Q}_m = \frac{Q_m}{1+ \frac{1}{2}Q_m^2},
\eeq
or, alternatively, 
\beq \label{Qm}
Q_m = \frac{1\pm\sqrt{1-2\alpha^2}}{\alpha}. 
\eeq

From \eqref{Qm}, we have that the reality of $Q_m$ implies that $\left|\alpha\right|$ is restricted to the domain $\left|\alpha\right| \in \left[0, \frac{1}{\sqrt{2}}\right]$. 
Furthermore, a given $\alpha$ has two associated $Q_m$ parameters, $Q_\pm$, in a self-explaining notation.
The $Q_+$ branch is restricted to the domain $\left|Q_+\right| \in [\sqrt{2}, \infty)$, with limits respectively attained at $\left|\alpha\right| = 1/\sqrt{2}$, $\alpha = 0$. The $Q_-$ branch satisfies $\left|Q_-\right| \in [0, \sqrt{2}]$, and its limits correspond to $\alpha=0, \left|\alpha\right| = 1/\sqrt{2}$. We observe that both branches merge at the critical value $\left|\alpha\right| =\alpha^* \equiv 1/\sqrt{2}$. 

The two-branch structure of \eqref{Qm} is particularly interesting.
Indeed, as we will show in the next subsection through an explicit computation, the $Q_-$ branch of magnetic solutions corresponds to the branch of  linearly stable pumping solutions, while the $Q_+$ branch maps onto the linearly unstable branch (see Section \ref{stab}). Remembering how the magnetic solution has been obtained performing the double Wick rotation \eqref{double_Wick} on the three-dimensional charged black brane solution \eqref{ds2_RN}, \eqref{f_RN}, \eqref{A_RN}, we bring into attention that $\left|Q_e\right| = \left|Q_m\right|$. Therefore, for $\left|Q_m\right|  = \left|Q_+\right| \geq \sqrt{2}$, the linear instability of this branch of pumping solutions seems to be related to the fact that our original $r^2 = 1 + R^2$ expansion is performed with respect to the wrong horizon (see comments below \eqref{temperature}). In the next Section we also show that $\alpha^*$, \emph{i.e.} the critical value where the two branches of \eqref{Qm} merge, corresponds to $\alpha_{b,{\rm max}}$.%
\footnote{We remind the reader that $\alpha_{b,{\rm max}}$ is the maximum value of $\alpha_b$ for which pumping solutions exist, see Section \ref{sps}.} This is an important and exact result emerging directly from the chain of mappings illustrated above.%
\footnote{It would be interesting to repeat the argument in higher dimensionality and get an exact value for the adiabatic threshold $\alpha_{b,{\rm max}}$ starting from an extremal charged AdS black hole in 3+1 dimensions.}

The different maps we have uncovered so far are not the end of the story. Another interesting geometry can be brought into the game. In \cite{Andrade:2013gsa}, Andrade and Withers (AW) introduced a beautifully simple holographic model of momentum relaxation. It involved charged black branes with nontrivial axion profiles along the spacelike boundary directions, that broke translational invariance at the level of the solution as a whole, but keeping the metric and the electromagnetic field translationally invariant. In three dimensions, the corresponding AW solution would involve a neutral massless scalar field $\phi(x) = \gamma x$. Focusing on the uncharged case, the metric of the AW solution would be given by 
\beqa
&&ds^2 = - f(r) dt^2 + f(r)^{-1} dr^2 + r^2 dx^2, \\
&&f(r) =  r^2 - M - \gamma^2 \log(r), \\
&&\phi(x) = \gamma x\ .
\eeqa
By Hodge duality, the scalar field is associated with the field strength 
\beq
F= \star d\phi = \gamma \star dx =\frac{\gamma}{r} dt \wedge dr, 
\eeq
that comes from the potential $A = -\gamma \log(r)dt$. If we identify $\gamma = Q_e$, we find that the charged black brane and the neutral three-dimensional AW solution are dual to each other. Of course, we must take $M = 1$  after the duality map to land on the one-parameter family of charged black branes \eqref{ds2_RN}, \eqref{f_RN} and \eqref{A_RN}. It is also possible to show that, by performing again the coordinate change $r^2 = 1 + R^2$ on the $M=1$ AW solution, double Wick rotating as $x \to i t, t \to i \varphi$, $\gamma \to - i \gamma$ and rescaling our coordinates, the pumping solution is obtained.

Let us summarize the structure of duality mappings through the following diagram,\footnote{We emphasize that the duality map between the electrically charged black hole and the magnetic solution has already been discussed in \cite{Cataldo:1996yr}.}%
$$
\begin{array}{ccc}
\hbox{charged black brane} & \overset{DW}{\xrightarrow{\hspace*{1cm}}}  & \hbox{magnetic AdS}_3 \\
\star\Big\downarrow & &\Big\downarrow   \star \\
\hbox{Andrade-Withers} & \overset{DW}{\xrightarrow{\hspace*{1cm}}}  & \hbox{pumping solution} \\
\end{array}
$$
where $DW$ stands for the double Wick rotation. 

\subsection{Analytic pumping solution in AdS$_3$}

As already anticipated before in Section \ref{commu}, we can employ the relation \eqref{r_R} to obtain an analytic expression 
for the three-dimensional pumping solution. Let us set $R = R(r)$, identify $\tilde{Q}_m = \alpha$ and recall that $\beta(\alpha) = 2\pi/(1+Q_\pm^2/2) = \pi (1\mp\sqrt{1-2\alpha^2})$. 
It turns out that \eqref{r_R} can be explicitly solved, yielding 
\beq
R(r)^2=-\frac{\beta(\alpha)^2}{4\pi^2} + \frac{1}{2}\alpha^2 W\left(\frac{\beta(\alpha)^2}{2\pi^2 \alpha^2} e^{\frac{2 r^2}{\alpha^2} + \frac{\beta(\alpha)^2}{2\pi^2\alpha^2}}\right), \label{R_r_analytic}
\eeq
where $W$ is the Lambert $W$-function, which solves the transcendental equation $W(x) e^{W(x)} = x$. Since $W(x>0)>0$, 
we consistently find that $R(r)^2 \geq 0$, with equality only at $r = 0$, a fact that is guaranteed by $W(x e^x) = x$. 
By taking \eqref{R_r_analytic} and substituting it in \eqref{h_R} and \eqref{d_R}, we obtain the three-dimensional pumping solution in a closed form. 
It can be straightforwardly checked that the resulting expressions for $f(r)$ and $\delta(r)$ solve the equations of motion of the pumping solution 
in Schwarzschild coordinates. Alternatively, if we take the relations \eqref{h_R} and \eqref{d_R} and introduce them into the equations of motion for the pumping solution with $R = R(r)$ unknown, we arrive to the equation 
\beq
R'(r) - \frac{r [\beta(\alpha)^2  + 4\pi^2 R(r)^2]}
{R(r) [2\pi^2\alpha^2 + \beta(\alpha)^2 + 4 \pi^2 R(r)^2]} = 0, 
\eeq
for which \eqref{R_r_analytic} is the only solution such that $R(r=0) = 0$.
\begin{figure}[h!]
\begin{center}
\includegraphics[width=8cm]{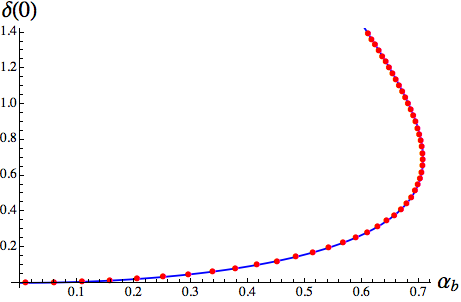}
\end{center}
\caption{\label{figure-3} \small $\delta(0)$ vs $\alpha$, numerical (red dots) and analytical (blue curve).}
\end{figure} 
\\
As a final consistency check, let us note that equation \eqref{d_R} implies that 
\beq 
\delta(0) = \log\left(1+\frac{1}{2}Q^2_\pm\right) =  \log\left(\frac{1\pm\sqrt{1-2\alpha^2}}{\alpha^2}\right). \label{delta_0} 
\eeq
In Fig.\ref{figure-3}, we plot $\delta(0)$ against $\alpha$,\footnote{Recall that $\alpha = \alpha_b$, since the map between the magnetic and the pumping solutions has been obtained in the boundary time gauge.} as obtained from the numerical solution of the equations of motion, and compare it with the expression \eqref{delta_0}. We plot both the stable and the unstable branches. The agreement is perfect: the numerically determined maximum of $\delta(0)$ for the stable branch is located at $\alpha = 0.7071067$, where $\delta(0) = 0.693148$. The expected results are $\alpha = 1/\sqrt{2} = 0.7071068$ and $\delta(0) = \log(2) = 0.693147$.

\section{Conclusions and future prospect}

Quasistatic adiabatic quenching procedures that interpolate between two different constant values for the external source can be modeled -for a long time interval- by a coupling constant that rises linearly with a fixed slope. In the case of a marginal deformation dual to a bulk massless scalar, 
the physics is in the meanwhile described by the pumping solution. We have explored the space of such solutions unravelling 
an interesting structure of stable and unstable branches. 

The pumping solutions are parametrized by the slope of the ramp $\alpha$ which can be expressed in either the boundary time gauge, $\alpha_b$, or the time gauge at the bulk center, $\alpha_o$. There is a range of $\alpha_o$ where the mass is negative, and this encompasses both stable and unstable solutions.
Only the stable branch is reachable by a physical quench from zero coupling and zero initial slope $\alpha_b(0)=0$. 
If one considers a ramp with a slope $\alpha_b(t)> \alpha_{b,\rm{max}}$, the system destabilizes and forms a black hole. 
This sets a limit to the physical adiabadicity. The breaking of adiabaticity is a known phenomenon in the 
vicinity of a critical phase transition  \cite{Das:2016eao}. It is intriguing to interpret the above 
limit as the presence of some quantum critical point that is crossed by the protocol. 
Regarding this fact, note that we have not analyzed the formation time $t_c$ of the resulting black hole once the adiabaticity threshold is surpassed. 
It would be interesting to elucidate if $t_c$ displays some universal scaling with respect to the deviation from the threshold $\alpha_{b,\rm{max}}$. 

Beyond the threshold value $\alpha_o\geq \alpha_{o,{\rm thr}}$ (where $\alpha_{o,{\rm thr}}$ corresponds to $\alpha_{b,\rm{max}}$ upon changing time gauge), linearized fluctuations become unstable. 
This is consistent with the fact that these solutions cannot be prepared through a quasistatic quench, but 
only exist as driving solutions that extend to $t\to -\infty$. The nonlinear analysis, however, unravels that 
for $\alpha_o\in (\alpha_{o,\rm{thr}}, \alpha_{o,c})$ the final fate of these unstable solutions is a limiting 
cycle rather than a black hole. Unfortunately, we have not been able to provide an alternative construction of the limiting cycle solutions, 
that would definitely establish their existence on firm grounds. This is another mandatory direction for future studies.

Among the numerous future extensions, one technical but interesting direction is the proper 
systematization of the holographic renormalization for this time-dependent circumstance. 
Another future direction sprouts from the observation that the model at hand is a special 
example of a wider class. Indeed one could consider complex, instead of real, massless bulk fields. 
The complex massless scalar fields still correspond to marginal deformations, but they allow for many intriguing features such as a non-trivial 
scalar radial profile and a natural connection to periodically driven systems. 
They also provide an intermediate step to study relevant deformations in a similar setup. 

We emphasize that the pumping solutions are driven setups and are not dual to vacuum configurations of the CFT. The Ward identity \eqref{mwi} and the vanishing of $\langle \cal{O}_\phi \rangle$ imply however that there is no exchange of energy through the non-trivial source. It would be thereby interesting to consider the relation of the pumping solutions with models of AdS cosmology.

Regarding the three-dimensional chain of maps presented in Section \ref{commu}, for the moment we prefer to view it as a purely formal construction. 
The question of whether this construction can lead to some useful and nontrivial physical insight is definitely 
interesting on its own right and will be discussed elsewhere.

\section*{Acknowledgments}

We would like to thank Andrea Amoretti, Daniel Are\'an, Riccardo Argurio, Anxo F. Biasi, Maciej Maliborski,  \'Angel Paredes,  Ioannis Papadimitriou, Andrzej Rosworowski, Alfonso Ramallo, Alberto P\'erez-Mu\~nuzurri and An\'\i bal Sierra-Garc\'\i a for pleasant, interesting and useful scientific exchanges.
The work of D.M. was supported by grants FPA2014-52218-P from Ministerio de Economia y Competitividad. A.S. is supported by the European Research Council grant HotLHC ERC-2011-StG-279579.  The work of J.M. and A.S. is partially supported by spanish grant PA2014-52218-P, by Xunta de Galicia (GRC2013- 024)  and by FEDER. This research has benefited from the use computational resources/services provided by the Galician Supercomputing Centre (CESGA).

\color{black}

\begin{appendix}

\section{The pumping solution from a magnetically sourced AdS$_4$}

The appearance of negative mass solutions is somewhat intriguing. In the literature, other similar cases have been shown to exist. The most famous case is the AdS-soliton which has the asymptotic spatial topology of $S^1\times {\mathbb R}^{d}$. Instead, our solution has the same asymptotic topology as usual AdS$_{d+1}$. It turns out that there is a mapping whereby the geometries can be seen to be equal to those of an Einstein gravity theory coupled with a $d-$form magnetic field strength. 
In this section we will make explicit this mapping  in the case of AdS$_4$, stressing that the construction can be made in any dimensionality. For the case of AdS$_3$, the dual magnetic solution has already been discussed in the main text.
In order to show the construction, we will move to standard radial coordinates $r = l \tan x$ and rewrite the metric ansatz \eqref{line1} as follows
\beq
ds^2 = - f(r) e^{-2 \delta(r)} dt^2 + f(r)^{-1} dr^2 + r^2 d\theta + r^2 \sin^2\theta d\varphi^2. \label{metric_ansatz}
\eeq
On top of this, let us switch a magnetic 2-form potential  
$
B_{r\varphi}(r,\theta) = 3r^2 \cos \theta V(r)\, 
$.
The field strength 3-form $F=dB$ only has one non-vanishing component
\be
F_{r\theta\varphi} = r^2 \sin\theta V(r)\, .
\ee
From the Maxwell equations $\partial_\alpha(\sqrt{-g} F^{\alpha\mu\nu}) = 0$  we get the following equation  for $V(r)$
\be
r fV' + (1 + 3r^2 +  f) V= 0\, .\label{Maxeq3}
\ee
The energy momentum tensor is  diagonal
\be
T_{\mu\nu} = {\rm diag} \left(  e^{-2\delta} f^2 , 1, r^2 f , r^2 f\sin^2\theta   \right) V^2\, ,
\ee
and with it we may compute the Einstein equations that lead to the following set
\beqa
f'  &=& \frac{1}{r} + 3r - \frac{f}{r} - \frac{f}{r} (1 + r^2 V^2) \\
\rule{0mm}{5mm}
\delta' &=& - r V^2
\eeqa
If we compare these equations to the ones for the adiabatic pumping ansatz  $\phi(t) = \alpha t$
\beqa
f' &=&  \frac{1}{r} + 3r - \frac{f}{r} \left(1 + r^2 \frac{\alpha^2 e^{2\delta(r)}}{f(r)^2} \right) \\
\delta'&=& -\frac{\alpha^2 e^{2\delta}}{f^2}
\eeqa
we can see that they are the same under the replacement 
\be
V(r) = \frac{\alpha e^{\delta(r)}}{f(r)}\, 
\ee
which, by the way, solves the Maxwell's equations \eqref{Maxeq3}.

\section{Constructing time-periodic pumping solutions}
\label{pumping_periodic}

In this appendix we provide a detailed discussion of the construction of the exactly periodic solutions described in Section \ref{sec_time_periodic}.

\subsection{Solving the perturbative ansatz}

We take the ansatz \eqref{periodic_perturbative_phi}-\eqref{periodic_perturbative_frequency} and impose the following boundary conditions: 
\begin{itemize}
\item Normalizability of the scalar field, $\varphi_{2n+1-k, k}(\tau, \pi/2) = 0$. 
\item Regularity of the scalar field at $x = 0$,  $\lim_{x\to0} x\, \varphi_{2n+1-k, k}(\tau, x) = 0$. 
\item Regularity of the blackening factor at at $x = 0$,  $\lim_{x\to0} x\, a_{2n-k, k}(\tau, x) = 0$. 
\item Preservation of the boundary time gauge, $d_{2n - k, k}(\tau, \pi/2) = 0$. 
\end{itemize}
Furthermore, and with no loss of generality, we also demand that $\phi(0, x) = 0$. Note that, when $\epsilon = 0$, the ansatz \eqref{periodic_perturbative_phi}-\eqref{periodic_perturbative_frequency} must reduce to the pumping solution. Therefore, we must have $\varphi_{2n+1, 0} = \partial_t a_{2n,0} = \partial_t d_{2n, 0} = 0$. On the other hand, for $\alpha_b = 0$, we must recover the perturbative expansion of the exactly periodic solution branching from the fundamental eigenmode of global AdS$_4$ \cite{Maliborski:2016zlh}.\footnote{The comparison of our results with \cite{Maliborski:2016zlh} is not immediate, since there the origin time gauge was employed.}
\\\\
Substituting \eqref{periodic_perturbative_phi}-\eqref{periodic_perturbative_frequency} into the equations of motion and expanding to order $n = 1$, we get the following equations for the metric  
\beqa
&&a_{2,0}' + \left(3 \tan x + \cot x \right)  a_{2,0} + \cos x\sin x = 0 ,\\
&&a_{1,1}' + \left(3 \tan x + \cot x \right)  a_{1,1} + 6 \sin x \cos^4\!x \cos \tau = 0,\\
&&a_{0,2}'+ \left(3 \tan x + \cot x \right)  a_{0,2} + \frac{9}{2} \sin x \cos^5\!x \left(1+\cos 2x\cos 2\tau \right) = 0, \\
&&d_{2,0}' + \cos x \sin x = 0, \\
&&d_{1,1}' + 6 \sin x \cos^4\!x \cos \tau = 0, \\
&&d_{0,2}' + \frac{9}{2} \sin x \cos^5\!x \left(1+\cos 2x \cos 2\tau \right) = 0, 
\eeqa
which are solved by 
\beqa
&&a_{2,0}(x) = \frac{1}{2}\cot^2\!x \sin 2x (x - \tan x), \\
&&a_{1,1}(\tau,x) = - \sin x \sin 2x \cos^2\!x \cos \tau,  \\
&&a_{0,2}(\tau, x) = -\frac{3}{2} \sin^2\!x \cos^6\!x \cos 2\tau + \frac{9}{64}\cot x\cos^2\!x(\sin 4x- 4x), \\
&&d_{2,0}(x) = \frac{1}{2}\cos^2\!x, \\
&&d_{1,1}(\tau, x) = \frac{6}{5}\cos^5\!x \cos \tau, \\
&&d_{0,2}(\tau, x) = \frac{3}{16} \cos^6\!x \left(4+ (3 \cos 2x- 1) \cos 2 \tau \right), 
\eeqa
where the aforementioned boundary conditions have been imposed. Regarding the scalar field, the equation for $\varphi_{3-k,k}$ takes the form 
\beqa
\varphi_{3-k,k}'' + 2 \csc(x) \sec(x) \varphi_{3-k,k}' - 9 \ddot{\varphi}_{3-k,k} + S_{3-k, k} = 0\ .
\eeqa
The differential operator acting on $\varphi_{3-k,k}$ is nothing but the AdS$_4$ Laplacian, 
expressed in $\tau = 3t$ and $x$ coordinates, and the non-homogeneous source $S_{3-k, k}$ depends on lower-order corrections. Explicitly, 
\beq
S_{3,0} = 0, 
\eeq 
\beq
\begin{split}
S_{2,1} =  &3 \cos^2\!x (2 \cos x (\omega_{2,0} + 3 (d_{2,0} - a_{2,0})) +\\
 &\sin x (d_{2,0}' - a_{2,0}'))\sin \tau  + 9 \cos^3\!x (\dot{a}_{2,0} - \dot{d}_{2,0})\cos \tau +  3(\dot{a}_{1,1} -\dot{d}_{1,1}), 
\end{split}
\eeq
\beq
\begin{split}
S_{1,2} =  &3 \cos^2\!x (2 \cos x (\omega_{1,1} + 3 (d_{1,1} - a_{1,1})) +\\
 &\sin x (d_{1,1}' - a_{1,1}'))\sin \tau  + 9 \cos^3\!x (\dot{a}_{1,1} - \dot{d}_{1,1})\cos \tau + 3(\dot{a}_{0,2} -\dot{d}_{0,2}) , 
\end{split}
\eeq
\beq
\begin{split}
S_{0,3} =  &3 \cos^2\!x (2 \cos x (\omega_{0,2} + 3 (d_{0,2}- a_{0,2})) +\\
 &\sin x (d_{0,2}' - a_{0,2}'))\sin \tau  + 9 \cos^3\!x (\dot{a}_{0,2} - \dot{d}_{0,2})\cos \tau. 
\end{split}
\eeq
When solving any one of the nontrivial equations above, the requirement that $\varphi_{3-k,k}$ 
is both regular at $x = 0$ and unsourced can only be satisfied if the frequency 
correction appearing in $S_{3-k,k}$ takes a particular value. We have that $\omega_{2,0} = -7/4$, $\omega_{1,1} = 0$ and $\omega_{0,2} = -135/128$. 
The most general solutions compatible with the boundary conditions that we find are:
\beq
\begin{split}
\varphi_{2,1}(\tau,x) = &\frac{3}{20}\cos^3\!x(19 \cos^2\!x + 5 x (2 (x + \cot x) + \sin 2x))\sin \tau +\\
&C_{2,1} \cos^3\!x \sin \tau, 
\end{split}
\eeq 
\beq
\begin{split}
\varphi_{1,2}(\tau,x) =  &-\frac{1}{17920}  \cos^2\!x \cot x (-3240 x + 5760 x \cos 2x - 3600 x \cos 4x  + \\
&1345 \sin 2x + 4394 \sin 4x + 381 \sin 6x ) \sin 2 \tau \\
&+C_{1,2}\cos^3\!x\sin \tau,  
\end{split}
\eeq 
\beq
\begin{split}
\varphi_{0,3}(\tau,x) = &\frac{3}{896}\cos^3\!x (108 \cos^4\!x - 88 \cos^6\!x + 12 \cos^8 \!x - \\
&252 x (x + 2 \cot x) + 63 \cos^2\!x (4x \cot x - 1) )\sin \tau - \\
&\frac{1}{64}\cos^9\! x(1+ 9 \cos 2x)\sin(3 \tau) + C_{0,3}\cos^3\!x \sin \tau + \\
&\frac{D_{0,3}}{64}\cos^3\!x (- 2 + 6 \cos 2x - 3 \cos 4x  + 4\cos 6x )\sin(3 \tau). 
\end{split}
\eeq 
We observe that regularity and normalizability by themselves are not enough to fix the undetermined constants $C_{1,2}, C_{2,1}, C_{0,3}, D_{0,3}$. Notice also that, up to this point, $\epsilon$ remains as a formal expansion parameter. By ascribing a physical meaning to it we can reduce the four undetermined integration constants to one. 
Let us choose the definition we employed in the main text and identify 
\beq
\left<\mathcal O_\phi (\tau = \pi/2)\right> = 3\epsilon. 
\eeq
Higher order corrections would modify this relation unless $C_{1,2}=0, C_{2,1}= - 3/8 \pi^2$ and $C_{0,3} = 3/128(9 \pi^2 - 10 D_{0,3})$. The integration constant $D_{0,3}$ cannot be fixed at this order of the perturbative expansion.\footnote{This phenomenon is solely due to the $\alpha_b = 0$ sector of the perturbative expansion and has been previously pointed out in  \cite{Maliborski:2016zlh}.} It turns out that, when computing the $\varphi_{0,5}$ correction, regularity and normalizability enforce that $D_{0,3} = 305/808$. 

The introduction of a finite pumping has nontrivial consequences regarding the Fourier decomposition of the exactly periodic solution in time. To wit, while in the $\alpha_b = 0$ case only odd multiples of the oscillation frequency $\Omega$ appear, for $\alpha_b \neq 0$ even multiples are also present, as exemplified by the $\varphi_{1,2}$ correction. It is mandatory to take this fact into account when designing a pseudospectral code able to find these solutions at the numerical level. 

\subsection{Numerical construction}

We start by decomposing our dynamical fields as
\beqa
&&\Pi(\tau,x) = \alpha_b+ \frac{3}{2}\alpha_b^3 \cos^2\!x + \cos^2\!x \hat{\Pi}(\tau,x),\\
&&\Phi(\tau, x) = \cos x \hat{\Phi}(\tau, x), 
\eeqa 
in such a way that $\hat\Pi, \hat\Phi$ satisfy the boundary conditions $\hat\Pi(\tau, \pi/2) = \hat\Phi(\tau, \pi/2) = 0$ 
with non-zero first-order spatial derivatives. The equations of motion are correspondingly modified. 
Taking into account the appearance of both odd and even multiples of the fundamental frequency at the perturbative level, 
we Fourier decompose our rescaled fields in time as 
\beq
\hat\Pi(\tau, x) = \sum_{k=0}^{N_k} \cos(k \tau) P_k(x), \qquad 
\hat\Phi(\tau, x) = \sum_{k=0}^{N_k} \sin( (k+1) \tau) Q_k(x),
\eeq
where we have allowed for a nontrivial zero-mode $P_0(x)$ in $\hat \Pi$; 
that would take care of the fact that, when $\epsilon =0$, $\Pi$
must reduce to field corresponding to the pumping solution at the given $\alpha_b$. $N_k$ is a numerical cutoff in the total mode number. 
The functions $\{P_k, Q_k, k =0...N_k\}$ must be further decomposed into a convenient spatial basis. 
A suitable choice was provided in \cite{ps} and exploited in-depth in \cite{Maliborski:2016zlh} in the spherically symmetric and sourceless case. For completeness, let us elaborate a bit on this choice. We largely follow \cite{Maliborski:2016zlh}, and we refer the reader to that useful reference for further information. 

First, let us take $y=\frac{2}{\pi}x$, in such a way that $y \in [0, 1]$;\footnote{In contrast with what done in the main text, as explained below \eqref{ext_grid}, we are here extending the domain of $x$ to include the boundary value $\pi/2$.}
then notice that, since origin regularity imposes a definite parity on the fields of our problem, we might extend them to the domain $y \in [-1, 0]$ in a univocal way. Define $\bar y \in [-1,1]$, and focus on the extended functions $\{\bar{P}_k(\bar y), \bar{Q}_k(\bar y), k =0...N_k\}$. We introduce Lagrange interpolating polynomials $\{l_j(\bar y), j=0...2 N_j+1\}$ that satisfy $l_j(\bar y_k) = \delta_{j,k}$ for some extended collocation grid $\{\bar y_j, j=0...2N_j+1\}$. We can write a polynomial approximation to any function $f : [-1,1] \rightarrow \mathbb R$ as   
\beq
\mathcal I_{2 N_j+1}f(\bar y) = \sum_{j=0}^{2N_j+1} f(\bar y_j) l_j(\bar y) \equiv \sum_{j=0}^{2N_j+1} \bar f_j l_j(\bar y). \label{int_approx}
\eeq
Explicitly, the interpolating polynomials are given by 
\beq
l_j(\bar y) =  \frac{w_j}{\bar y- \bar y_j} \left( \sum_{l=0}^{2N_j+1} 
\frac{w_l}{\bar y- \bar y_l} \right)^{-1} ,
\eeq
where the weights $\{w_j, j=0...2N_j+1\}$ are defined as $w_j = \left(\prod_{j=0,j\neq l}^{2N_j+1} (\bar y_j - \bar y_l) \ \right)^{-1}$ in terms of the extended collocation grid. So far, the extended collocation grid we are referring to remains arbitrary. 
A convenient choice is provided by the Chebyshev collocation grid of the second kind 
\beq
\bar y_j = \cos\left(\frac{\pi j}{2 N_j+1} \right), j=0...2N_j +1. \label{ext_grid}
\eeq
Notice that the boundary $x=\pi/2$ (\emph{i.e.} $\bar y=1$) is included in the collocation grid, while the origin is avoided. This last feature is convenient in two regards. First, it allows to impose the boundary conditions at $x = \pi/2$ on the functions $\{\bar{P}_k(\bar y), \bar{Q}_k(\bar y), k =0...N_k\}$ in a straightforward way. Second, it implies that the singular behavior of some terms present in the equations of motion at $x = 0$ is no longer a concern in the discretized version of the problem. Furthermore, with this choice, the approximant \eqref{int_approx} satisfies 
\beq
I_{2N_j+1}f(\bar y) = \sum_{j=0}^{2 N_j+1} \hat f_j T_j(\bar y), 
\eeq
where $\{T_j(\bar y), j = 0...2N_j+1\}$ are Chebyshev polynomials of the first kind. Standard optimization theorems in polynomial approximation then apply immediately to our case. As a final benefit, it turns out that the grid choice \eqref{ext_grid} allows for a simple analytic expression of the $w_j$ weights \cite{ps}. The $n$-th derivative of the function $f$ can be approximated by the $n$-th derivative of \eqref{int_approx}. We have that 
\beq
\frac{d^n}{d\bar y^n} \mathcal I_{2Nj+1}(\bar y_j) = \sum_{l=0}^{2Nj+1} D_{j,l}^{(n)} \bar f_l\ ,
\eeq
where the explicit form of the $2(N_j+1) \times 2(N_j+1)$ differentiation matrix $\bold{D}^{(n)}$ can be found in \cite{Maliborski:2016zlh}. Now the crucial point comes into play. Since we are working with functions of definite parity around $x = 0$, $P_k(-x) = P_k(x), Q_k(-x)=-Q_k(x)$, we have that, being $f$ any one of these functions and $p$ its parity under $x \to -x$, 
\beq
\bar f_{2 N_j +1 -j} = (-1)^p \bar f_j \equiv (-1)^p f_j, j=0...N_j\ ,
\eeq
in such a way that the differentiation matrix $\bold{D}^{(n)}$ splits into four $(N_j +1) \times (N_j +1)$ blocks. Defining $\bold f = (f_0,...,f_{N_j})$, $\bold{\bar f}= (f_0,..., f_{N_j}, (-1)^{p} f_{N_j}, ..., (-1)^p f_0)$ we have that 
\beq
\bold{\bar f} = \begin{pmatrix}
\bold 1 & \bold 0\\
(-1)^p \bold R & \bold 0
\end{pmatrix} \begin{pmatrix} \bold f \\ \bold 0 \end{pmatrix},
\eeq
where $\bold R$ is a $(N_j+1) \times (N_j +1)$ exchange matrix. In consequence  
\beq
\bold{D}^{(n)}\bold{\bar f} = \begin{pmatrix}
\bold {D}_{++}^{(n)} & \bold{D}_{+-}^{(n)}\\
\bold{D}_{-+}^{(n)} & \bold{D}_{--}^{(n)}
\end{pmatrix} \begin{pmatrix}
\bold 1 & \bold 0\\
(-1)^p \bold R & \bold 0
\end{pmatrix} \begin{pmatrix} \bold f \\ \bold 0 \end{pmatrix} = 
\begin{pmatrix} (\bold{D}_{++}^{(n)} + (-1)^p \bold{D}_{+-}^{(n)}\bold R) \bold f \\ (\bold{D}_{-+}^{(n)} + (-1)^p \bold{D}_{--}^{(n)}\bold R) \bold f \end{pmatrix}, 
\eeq
and we find that the parity-adapted $n$-th derivative operator acting on the physical $y > 0$ part of the extended grid is  
\beq
\bold{D}^{(n)}_p \equiv \bold{D}_{++}^{(n)} + (-1)^p \bold{D}_{+-}^{(n)}\bold R \ . 
\eeq
Therefore, for functions of well-defined parity around $x=0$, if we restrict ourselves to the original spatial domain $x \in [0, \pi/2]$ but employ the rescaled collocation grid 
\beq
x_j = \frac{\pi}{2}\cos\left(\frac{\pi j}{2 N_j+1}\right), j=0...N_j \ ,
\eeq
and discretize derivatives with the just defined operator $\bold{D}^{(n)}_p$ , the desirable features of working with a Chebyshev spectral decomposition are kept while also the boundary conditions at $x = 0$ are automatically incorporated into the discretized problem.

Having clarified our strategy, we take the final ansatz 
\beqa
&&\hat\Pi(\tau, x_j) = \sum_{k=0}^{N_k} \sum_{j=0}^{N_j}\cos(k \tau) p_{k,j}\ , \\
&&\hat\Phi(\tau, x_j) = \sum_{k=0}^{N_k}\sum_{j=0}^{N_j} \sin( (k+1) \tau) q_{k,j}\ ,
\eeqa
so we work in mode space in $\tau$, but real space in $x$. As a collocation grid in the time domain, we choose 
\beq
\tau_k = \frac{2 \pi (k-1/2)}{2 N_k + 3}, k=1...N_{k}+1\ ,
\eeq
At each $\tau_k$, it is convenient to define the variables $\{\delta_{k,j},c_{k,j}\}$, 
where $\delta_{k,j} = \delta(\tau_k, x_j)$ and $c_{k,j} = f(\tau_k, x_j) \exp(-\delta(\tau_k, x_j))$, 
which are then expressed in terms of $\{p_{k,j}, q_{k,j}\}$ by discretizing the corresponding constraint 
equations and inverting the resulting discretized differential operators. 
Imposing the boundary conditions $\delta_{k,0}=0$ and $c_{k,0}=1$ ensures that we are working in the boundary 
time gauge and renders the above mentioned discretized differential operators invertible. Once this expressions are known, the first-order dynamical equations for the scalar field can be solved by means of a Newton-Raphson algorithm that we implement in Mathematica. We impose the boundary conditions $p_{k,0} = q_{k,0} = 0$, that follow from the linear independence of the Fourier modes of the time decomposition and  $\hat\Pi(\tau, \pi/2) = \hat\Phi(\tau, \pi/2) = 0$. 

As a last comment, notice that we have $2 (N_k +1) \times (N_j +1) + 1$ dynamical variables, that correspond to $\{p_{k,j}, q_{k,j}\}$ and the oscillation frequency $\Omega$, but the discretization of the equations of motion for $\hat \Pi, \hat \Phi$ on the two-dimensional collocation grid provides just $2 (N_k +1) \times (N_j +1)$ equations. The remaining equation comes from normalizing some relevant physical quantity to a prescribed numerical value. 
We choose to set 
\beq
\left<\mathcal O_\phi(\pi/2)\right> = \lambda\ , 
\eeq
where $\lambda$ is an user-defined value for the vev amplitude. Since we have defined $\left< \mathcal O_\phi(\tau)\right> = -1/2\phi'''(\tau, \pi/2) = \hat\Phi''(\tau, \pi/2)$, this extra boundary condition is discretized as  
\beq
\sum_{k=0}^{N_k} \sin((k+1)\pi/2) \sum_{l=0}^{N_j} D^{(2)}_{0, l}q_{k,l} = \lambda\ .
\eeq
Finally, a given family of time-periodic solutions is found iteratively: 
we place ourselves at $\lambda, \alpha_b \ll1$ and employ the perturbative solution as an initial 
seed to start the relaxation algorithm. Then, we move in discrete steps along the $(\alpha_b(\eta), \lambda(\eta))$ curve in the two-dimensional 
phase space of time-periodic solutions, taking the one found in the previous step as the initial guess for the next step. 
The results shown in the main text of the paper have been obtained for an $N_k = N_j = 20$ grid; the time it took to find each solution along the curve was $O(5)$ min.  

\section{The relative entropy and the adiabatic quench}

Consider two density matrix operators, $\rho$ and $\sigma$, that describe two states of a quantum-mechanical system with Hilbert space $\mathcal H$. Their relative entropy
\beq
S(\rho||\sigma) = \textrm{tr}(\rho \log \rho) - \textrm{tr}(\rho \log \sigma) \ ,
\eeq
provides a measure of their distinguishability. In particular, $S(\rho||\sigma) \geq 0$ by construction, where equality holds only for $\rho = \sigma$. We shall refer to $\sigma$ as the reference state. Given that $\sigma$ is Hermitean and positive definite, it can be expressed as 
\beq
\sigma = e^{-H_\sigma}\ ,
\eeq
where the Hermitean operator $H_\sigma$ is known as the modular Hamiltonian. In terms of $H_\sigma$, the relative entropy reads
\beq
S(\rho||\sigma) = \Delta \left<H_\sigma \right> - \Delta S \ , \label{rel_ent_2}
\eeq 
where $\Delta \left<H_\sigma \right> = \left<H_\sigma \right>_\rho - \left<H_\sigma \right>_\sigma$ and $\Delta S = S_\rho - S_\sigma$, with $S$ being the von Neumann entropy of the associated density matrix (see for example \cite{VanRaamsdonk:2016exw} for an extensive review with references).

For a generic $\sigma$, the modular Hamiltonian is a nonlocal operator whose explicit form is not known. The situation changes only in a handful of cases, where the modular flow associated to $H_\sigma$, generated by the unitary operator $U_\sigma(s) = e^{-i s H_\sigma}$, admits a geometric interpretation. Consider a CFT placed on the Einstein Static Universe $ESU_{d+1} = \mathbb R \times S^d$, and take the cap-like region $C = \{t=t_0, \theta < \theta_0 : x\in ESU_{d+1}\}$. Let the theory be in its vacuum state $\left| 0 \right>$, and consider the reduced density matrix $\rho_\mathcal{C}$, which encodes any observation restricted to the causal development of $C$, $\mathcal C \equiv D(C)$. This reduced density matrix is obtained after partial trace over the complementary of $C$, 
$\rho_\mathcal{C} \equiv \textrm{tr}_{\bar C} \left|0\right> \left<0\right|$. The modular Hamiltonian $H_\mathcal C \equiv - \log \rho_\mathcal{C}$ is, in this case, a local operator given by 
\beq
H_\mathcal{C} = 2 \pi \int_{\theta < \theta_0} \frac{\cos\theta- \cos\theta_0}{\sin \theta_0}\ \hat{T}_{00} \ ,  \label{Hmod_cap}
\eeq
where $\hat{T}_{\mu\nu}$ is the CFT energy-momentum tensor operator. 

\smallskip

Recalling \eqref{rel_ent_2}, we observe that the AdS/CFT correspondence allows for an explicit computation of $S(\rho||\sigma)$ if we take $\rho_\mathcal{C}$ as reference state. Consider a fixed CFT, and a generic state $\rho$ with a dual geometric description. The GKPW prescription and the holographic renormalization procedure allow us to compute
\beq \label{TT}
\Delta\left<H_\mathcal{C}\right> =  2 \pi \int_{\theta < \theta_0} \frac{\cos\theta- \cos\theta_0}{\sin \theta_0} \left(\langle{\hat T}_{00}\rangle_\rho - \langle{\hat T}_{00}\rangle_{\rm vac}\right) \ .
\eeq
The CFT vacuum state is now described geometrically by global AdS$_{d+1}$. On the other hand, the holographic entanglement entropy (HEE) prescription provides a way of computing the difference of von Neumann entropies $\Delta S$, since now $S_\rho - S_\textrm{vac}$ is nothing but the area of an extremal surface homologous to $C$ regularized with respect to the vacuum geometry. Positivity of the relative entropy implies then that the inequality
\beq
\Delta\left<H_\mathcal{C}\right>  \geq \Delta S \ , \label{S_rel_ineq}
\eeq
must be satisfied if the non-vacuum geometry under question allows for a consistent quantum-mechanical interpretation as the holographic dual of the state $\rho$. 

A direct holographic computation where the vev's of the two energy-momentum in \eqref{TT} are evaluated explicitly considering the subleading mode of the dual metric, shows that for the pumping solution and for any opening angle $\theta_0$ of the cap, $\Delta\left<H_\mathcal{C}\right> \leq 0$. On the other hand, it is possible to show that $\Delta S \geq 0$, both by a perturbative computation at $\alpha_b \ll 1$ as well as numerically for finite $\alpha_b$. Hence, the negative energy density associated to the pumping solution rises an apparent violation of the above formula. Note however that the holographic computation needs probably to be reconsidered and refined. Indeed, we have to deal properly with the introduction of the background source $\phi_0 = \alpha_b t$. Let us linger on this important point more extensively. The deformation of our original CFT by means of the new time-dependent coupling implements an exactly marginal flow in theory space. As a consequence, the straightforward application of the GPKW prescription\footnote{$W_\textrm{ren}$ is the connected generating functional for the boundary theory.}
\beq
W_{ren}(\gamma_{\mu\nu}, \phi_0) = S_{ren}(\gamma_{\mu\nu}, \phi_0)\ , \label{basic}
\eeq
to the pumping solution would allow us only to compute the vev of the stress-energy operator defined {\it with the linearly rising scalar source}. While the stress-energy operators appearing in \eqref{TT} are both defined with the scalar source set to a constant value.

In other words, consider the holographic quench that produces the pumping solution at $t \rightarrow \infty$. This corresponds to the time-dependent deformation introduced by the background field $\phi_0(t)$, such that $\phi_0(t\leq0)=0$, $\lim_{t\to\infty} \phi_0(t)/t = \alpha_b$. At a generic $t > 0$, the relation \eqref{basic} would be computing the expectation value of the energy-momentum tensor of the deformed theory, and not of our original theory. These two operators, which we choose to denote respectively as $\hat{T}^{[\phi_0]}_{\mu\nu}$ and $\hat{T}^{[\phi_0=0]}_{\mu\nu}$, 
are generically {\it different}, because of the nontrival deformation of our initial CFT due to the marginal coupling. Relation \eqref{basic} guarantees then access to $\langle\hat{T}^{[\phi_0]}_{\mu\nu}\rangle_t$, but {\it not} to $\langle\hat{T}^{[\phi_0=0]}_{\mu\nu}\rangle_t$. Therefore the procedure we outlined to compute $S(\rho||\sigma)$ holographically cannot in general be applied directly to our case.
We would need additional information to know precisely how the two operators $\hat{T}^{[\phi_0=0]}_{\mu\nu}$ and $\hat{T}^{[\phi_0]}_{\mu\nu}$ are related by the time-dependent deformation; for instance, access to the microscopic description of the dual CFT would clarify the point.

It is interesting to note that we can invert the logical flow and use the potential paradox 
implied by violating \eqref{S_rel_ineq} to guess some features of the deformed theory. 
Suppose the deformed and undeformed energy-momentum operators satisfy a relation of the form%
\footnote{For our purposes, however, it is only important that whatever is to be put on the right hand side has vanishing vev.
Having a more articulated rhs is in fact a possible way out of the argument described in the main text.} 
\beq
\hat{T}^{[\phi_0]}_{\mu\nu} - \hat{T}^{[\phi_0=0]}_{\mu\nu} \propto \phi_0(t) \gamma_{\mu \nu} \mathcal O_\phi \ , \label{operator_relation}
\eeq
In the pumping solution, at sufficiently late values of time, $\langle\mathcal O_\phi \rangle_t = 0$ and hence we would have
\beq 
\langle\hat{T}^{[\phi_0]}_{\mu\nu}\rangle_t = \langle\hat{T}^{[\phi_0=0]}_{\mu\nu}\rangle_t \ .\label{paradox_eq}
\eeq
In such a case, the above-mentioned holographic computation would give the correct result and the lhs of equation \eqref{S_rel_ineq} would be negative, violating  the entropy inequality. At this point, we refrain from reaching provocative conclusions, given the uncertainties that we have spelled out concerning the correct way of computing the left hand side of \eqref{S_rel_ineq}. 

We would like to mention at this point the reference \cite{OBannon:2016exv} that appeared the same day as ours in which they consider a similar marginal scalar sourced linearly in time.
They also find a discrepancy in the so called Firt Law of Entanglement Entropy, which is \eqref{S_rel_ineq} in the limit of infinitesimal deformation. Altogether this points out an interesting
direction of future research.

\end{appendix}


\end{document}